\documentclass[journal,twoside,web]{ieeecolor}
\usepackage{tmi}
\usepackage{cite}
\usepackage{amsmath,amssymb,amsfonts}
\usepackage{algorithmic}
\usepackage{graphicx}
\usepackage{textcomp}
\usepackage{relsize}
\usepackage{soul}
\usepackage{subcaption}
\usepackage{booktabs}
\usepackage{multirow}
\usepackage[ruled]{algorithm2e}
\usepackage{array}
\usepackage{hyperref}
\graphicspath{{figures/}}

\def\BibTeX{{\rm B\kern-.05em{\sc i\kern-.025em b}\kern-.08em
    T\kern-.1667em\lower.7ex\hbox{E}\kern-.125emX}}
\begin{document}
\title{Topological Similarity Index and Loss Function for Blood Vessel Segmentation}
\author{Ricardo J. Ara{\'u}jo, Jaime S. Cardoso, \IEEEmembership{Senior Member, IEEE}, and H{\'e}lder P. Oliveira
\thanks{This work was financed by National Funds through the Portuguese funding agency, FCT - Funda{\c c}{\~a}o para a Ci{\^e}ncia e a Tecnologia within PhD grant number SFRH/BD/126224/2016.}
\thanks{R. J. Ara{\'u}jo is with INESC TEC, Porto, Portugal (e-mail: ricardo.j.araujo@inesctec.pt). }
\thanks{J. S. Cardoso is with INESC TEC and the Faculty of Engineering of the Univeristy of Porto, Porto, Portugal (e-mail: jaime.cardoso@inesctec.pt).}
\thanks{H. P. Oliveira is with INESC TEC and the Faculty of Sciences of the University of Porto, Porto, Portugal (e-mail: helder.f.oliveira@inesctec.pt).}}

\maketitle

\begin{abstract}
Blood vessel segmentation is one of the most studied topics in computer vision, due to its relevance in daily clinical practice. Despite the evolution the field has been facing, especially after the dawn of deep learning, important challenges are still not solved. One of them concerns the consistency of the topological properties of the vascular trees, given that the best performing methodologies do not directly penalize mistakes such as broken segments and end up producing predictions with disconnected trees. This is particularly relevant in graph-like structures, such as blood vessel trees, given that it puts at risk the characterization steps that follow the segmentation task. In this paper, we propose a similarity index which captures the topological consistency of the predicted segmentations having as reference the ground truth. We also design a novel loss function based on the morphological closing operator and show how it allows to learn deep neural network models which produce more topologically coherent masks. Our experiments target well known retinal benchmarks and a coronary angiogram database.
\end{abstract}


\section{Introduction}
\label{sec:introduction}
\IEEEPARstart{B}{lood} vessels are analysed every single day in clinical practice given their association to several pathologies with a high impact in our lives. Cardiovascular diseases such as strokes and heart attacks are the major cause of deaths worldwide~\cite{a1}, and an early detection is crucial to reduce the mortality~\cite{a2}. Another frequent complication due to blood vessel damage is diabetic retinopathy, a medical condition resulting from the effect of unregulated sugar levels in the blood and its impact in the retinal blood vessels. If left untreated, its progression is one of the major causes of vision loss worldwide~\cite{a3}. The analysis of blood vessels is also crucial in cancer therapeutics~\cite{a4,a5}, and surgery eligibility and planning studies~\cite{a6,a7,a8}.

Given the frequency and impact of blood vessel related conditions, there is a large volume of data generated every day. The number of experts is scarce, especially in low-to-middle income countries, thus the introduction of computer vision routines in the clinical pipelines for supporting the analysis of blood vessels is of utmost importance. Then, with little surprise, the topic of blood vessel segmentation in medical images was already being explored by the computer vision community at the 1980s decade~\cite{a9,a10}. For the following years, the trend continued to be encoding the prior knowledge on the structure of vascular networks in algorithms such as matched filtering~\cite{a10,a11,a12}, centreline tracking~\cite{a9,a13,a14,a15}, mathematical morphology~\cite{a16,a17}, and Hessian-based enhancement~\cite{a18,a19,a20}. With the evolution of computer hardware, and the emergence of publicly available datasets, machine learning also started to be explored~\cite{a21,a22,a23}. The most focused scenario was retinal blood vessel segmentation, possibly because the only datasets having complete manual annotations of the blood vessels were those containing retinal fundus photos. Nonetheless, many authors continued to explore more traditional approaches for coronary~\cite{a24}, lung~\cite{a25}, and brain~\cite{a26} vascular tree segmentation. Recently, similarly to what may be observed throughout most fields of computer vision where significant amounts of data are available, deep learning has reached new performance levels in this scenario~\cite{a27,a28,a29}.

\begin{figure*}[t!]
	\begin{center}
		\includegraphics[width=0.7\textwidth]{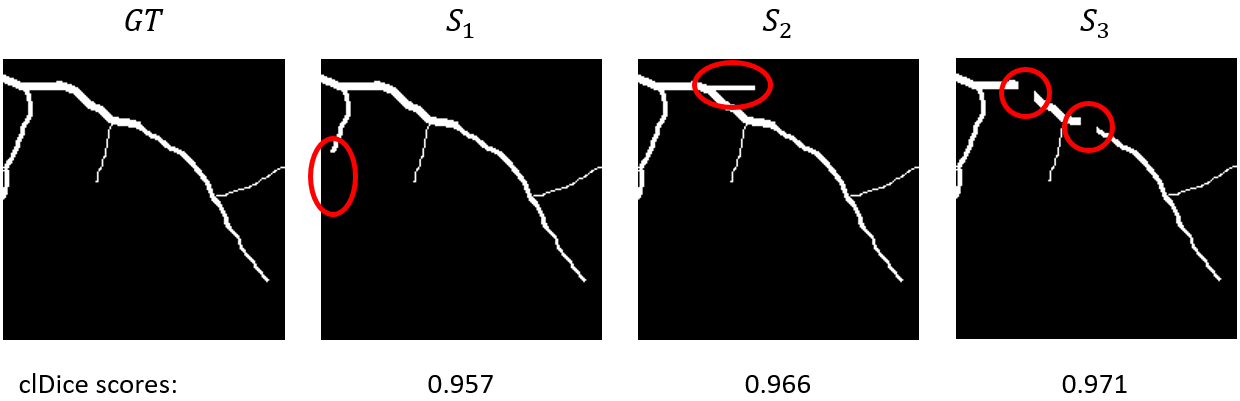}
		
		\caption{The effect of different errors on the clDice metric. $GT$ is the ground-truth mask and $S_1, S_2,$ and $S_3$ are segmentations where, respectively, a termination segment is missing, a false positive branch is added, and two broken segments are present (errors highlighted in red). As shown, the clDice metric is sensitive to the extension of centreline error, not to the effect of the errors in the overall graph. A topological metric should penalize $S_3$ more than $S_1$ and $S_2$.}
		\label{fig:cldice_invariance}
	\end{center}
\end{figure*}

Despite the success of deep learning methodologies, it is still challenging to obtain topological coherent masks, as they usually contain errors (such as broken segments) which have a large impact in the vascular tree graph and are likely to put at risk any automated characterization step that follows. This behaviour is expected since the Binary Cross Entropy (BCE) loss function is the usual choice for optimizing the model parameters and it does not properly highlight this kind of errors. Moreover, the topological coherence of predictions is often overlooked as the metrics that are usually reported (accuracy, sensitivity, specificity, and the Area Under the receiver-operating characteristic Curve (AUC)) are all based on the pixel-level and cannot clearly express such errors. Recently, this issue has been alleviated with the introduction of neural network designs based on probabilistic~\cite{a30} and recursive~\cite{a31} refinement models. In these works, the topological properties of the algorithms were assessed by sampling paths from the ground truth mask and determining the proportion of infeasible (impossible to reach the end point from the initial one), wrong (a path exists but is not equivalent to the ground truth one) and correct paths in the predicted mask. Even then, it may not always be trivial to objectively compare two different algorithms, as a decrease in the amount of infeasible paths will be likely followed by an increase in both wrong and correct paths, which naturally constitutes a compromise, similarly to sensitivity and specificity. Moreover, these metrics do not fully capture the amount of false positive paths being introduced by the algorithms, as any new sub-tree that does not exist in the ground truth will not be accounted for. A novel similarity metric derived from the Dice score and focusing the blood vessel skeleton has been proposed~\cite{a32}. The authors reported that the soft-clDice, the loss derived from the introduced metric, improves the connectivity of the final vascular trees. This is a natural outcome of the additional attention that is being given to the blood vessel centrelines, which themselves represent the vascular graph. However, we argue that the clDice metric is not a truly topological metric, since it yields the same penalty for a missing vessel termination and a broken segment of similar extension, when the latter has a much larger impact on the vascular tree graph. This is illustrated in Fig.\ref{fig:cldice_invariance}. Therefore, we believe the literature is lacking a unified metric that objectively represents these topological properties in a single value and a topological loss which can be used when training neural network models.

\subsection{Main Contributions}

The main contributions of this work are two-fold:

\begin{itemize}
	\item we propose a similarity index for blood vessel segmentation that assesses the topological coherence between the ground truth and predicted masks, penalizing more errors having a higher impact in the vascular tree graph; this behaviour is achieved by noticing that the impact of an error on the graph is related to the amount of paths it affects;
	
	\item a loss function penalizing false negative segments that induce disjoint trees and false positive segments which merge distinct trees is shown to be useful for learning models that perform better topological-wise. Our experiments cover both retinal and coronary blood vessel segmentation.
\end{itemize}

\subsection{Document Structure}

This Section briefly discussed the relevance of blood vessel segmentation in clinical practice and the evolution that its literature has been facing. The motivations for having performance indicators which are able to show the topological coherence of the segmentations were stated.
The rest of this document is structured as follows: Section 2 describes the properties that a topological metric should have and proceeds to the proposal of our topological similarity index; in Section 3 we design a loss function based on the morphological closing operator which allows to learn deep networks that are superior topological-wise; Section 4 describes the experiments we conducted and the respective findings; and finally, Chapter 5 concludes the paper and provides some directions for further research on this topic.

\section{A topological benchmark} \label{sec:2}

Blood vessel trees are graph-like structures in the sense that, besides local caliber, they are well encoded by a graph $G = (V, E)$, where vertices $V$ represent bifurcations and vessel terminations, and undirected edges $E$ encode the segments connecting them. In this paper, we designate by topological coherence the similarity between two vascular tree graphs, such as the ones corresponding to the ground-truth and predicted segmentations of a given image.

\subsection{Desired properties}

We start by stating the properties we believe a topological benchmark should possess: 

\begin{itemize}
	\item \textbf{Property 1}: to clearly highlight the errors that have impact on the vascular tree graph, such as broken and missing segments;
	\item \textbf{Property 2}: to further penalize broken segments since this type of errors is responsible for major changes of the underlying graph;
	\item \textbf{Property 3}: topological errors in the main vascular tree branches should have larger impact, since they may lead to large sub-trees being lost in automated analysis algorithms.
\end{itemize}

The most commonly used metrics for evaluating blood vessel segmentation algorithms - accuracy, sensitivity, specificity, and AUC - do not possess any of these properties. Even though they penalize broken and missing segments, these errors tend to be scarcer than caliber-related ones, thus they are strongly dissipated. The clDice metric accounts for the first property, since it focuses centerlines and neglects the errors due to caliber assessment. Nevertheless, as shown in Fig.~\ref{fig:cldice_invariance}, it equally penalizes broken and missing termination segments of similar extension, and also does not distinguish errors occurring in major and narrow branches. Therefore, to the best of our knowledge, there is no metric in the blood vessel literature providing properties 2 and 3.

\subsection{Topological similarity index}

Let $\mathbf{A}$ be a binary blood vessel mask, which can consist of a single connected component or multiple sub-trees. Let $\mathbf{A}_f$ be the set of vessel (foreground) pixels and $P_{i,j} = \{{\mathbf{x}_i,\ldots,\mathbf{x}_j}\}$ be the minimum cost path between pixels $i,j\in \mathbf{A}_f, l_i=l_j$, where $l_k$ specifies the sub-tree to which a vessel pixel $\mathbf{x}_k$ belongs.

\begin{figure}[t]
	\begin{center}
		\begin{minipage}{0.75\columnwidth}
			\begin{subfigure}{0.46\textwidth}
				\includegraphics[width=1\textwidth]{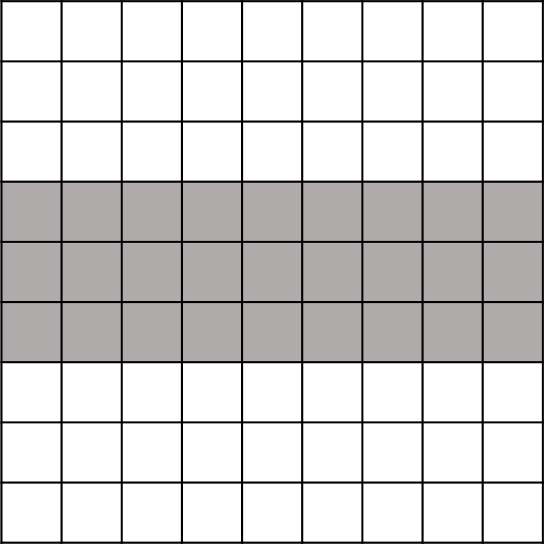}
				\caption{}
				\label{fig:ms_pa}
			\end{subfigure} \;
			\begin{subfigure}{0.46\textwidth}
				\includegraphics[width=1\textwidth]{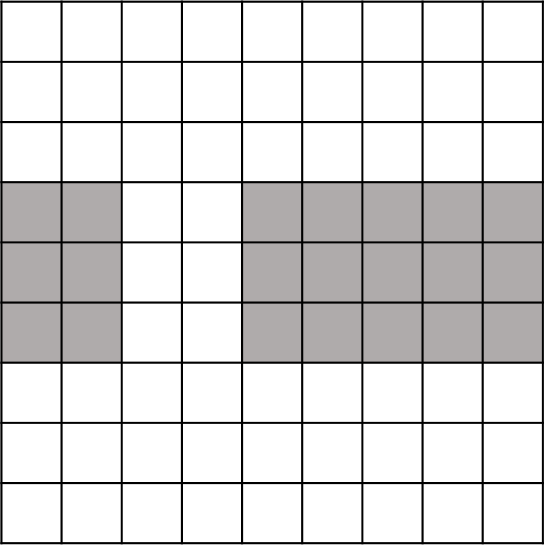}
				\caption{}
				\label{fig:ms_p_d1a}
			\end{subfigure}
			
			\begin{subfigure}{0.46\textwidth}
				\includegraphics[width=1\textwidth]{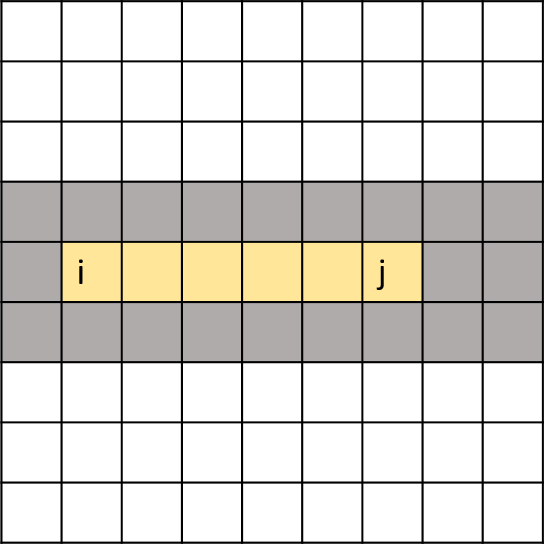}
				\caption{}
				\label{fig:ms_p_c1a}
			\end{subfigure} \;
			\begin{subfigure}{0.46\textwidth}
				\includegraphics[width=1\textwidth]{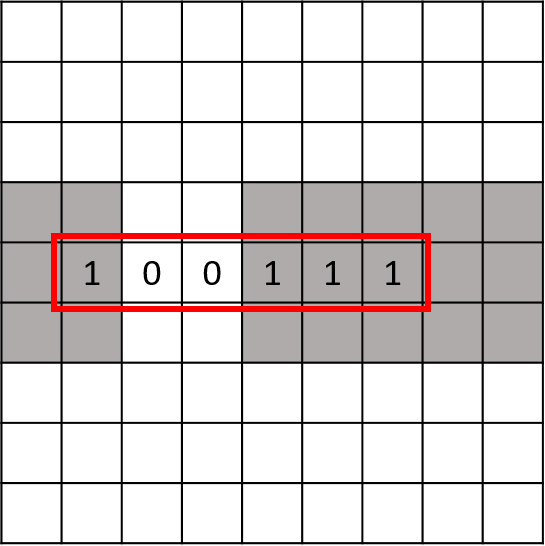}
				\caption{}
				\label{fig:ms_ya}
			\end{subfigure}
		\end{minipage}
	
		\begin{subfigure}{0.3\columnwidth}
			\includegraphics[width=1\textwidth]{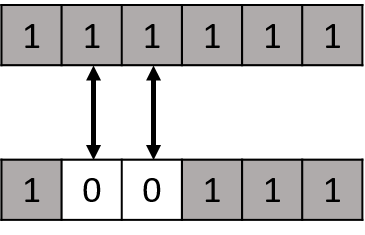}
			\caption{}
			\label{fig:ms_p_d2a}
		\end{subfigure}		
	\end{center}
	\caption[Hamming distance between a path sampled from a mask, and another mask.]{Hamming distance between a path $P_{i,j}$ sampled from a mask $\mathbf{A}_1$, and a mask $\mathbf{A}_2$. (a) $\mathbf{A}_1$, (b) $\mathbf{A}_2$, (c) a path $P_{i,j}$ sampled from $\mathbf{A}_1$, (d) the respective string obtained from $\mathbf{A}_2$, and (e) the number of switches needed to have both strings match, in this case, $H(P_{i,j}, \mathbf{A}_2) = 2$, therefore $f_H = 4/6$.}
	\label{fig:hamming}
\end{figure}

Removing a termination blood vessel segment $T=\{ \mathbf{x}_{t_1}, \ldots, \mathbf{x}_{t_n} \}$ would render impossible all paths $P_{i,j}$ where $\mathbf{x}_{i} \in T \lor \mathbf{x}_j \in T$. Considering a broken segment $B=\{ \mathbf{x}_{b_1}, \ldots, \mathbf{x}_{b_n} \}$ instead, in addition to paths $P_{i,j}$ where $\mathbf{x}_{i} \in B \lor \mathbf{x}_j \in B$, those where $\exists \mathbf{x}_i,\ldots,\mathbf{x}_j \in B$ would also not be possible anymore. Note how this relates with the second property we have specified, as a broken segment has a larger impact than a missing termination in terms of the amount of paths that become impossible (assuming errors of similar size). In fact, it is also likely to satisfy the third property due to the pattern naturally displayed by blood vessel trees. Given their root, they keep dividing in branches, and calibre decreases with each such division. Therefore, there is a natural tendency for the majority of possible paths in a vascular tree to traverse the major branches. This behaviour makes us consider functions taking into account the feasibility of paths when designing our proposed topological similarity index. Based on this motivation, we now formulate a similarity index $m: \mathbb{R}^D \times \mathbb{R}^D \rightarrow \left[0, 1\right]$ comparing two D-dimensional binary masks. We are interested in assessing how feasible the possible paths in the ground truth $\mathbf{Y}$ are in a segmentation $\mathbf{P}$ (path sensitivity/recall) and also the amount of false positive paths that exist in $\mathbf{P}$ (path precision). Let $N_P$ and $N_Y$ denote, respectively, the number of possible paths in $\mathbf{P}$ and $\mathbf{Y}$. An expression for a similarity index taking into account what was discussed until now follows:

\begin{equation} \label{eq:metric_general3}
\begin{split}
m&(\mathbf{P},\mathbf{Y}) = \\
& \sqrt{\frac{\sum\limits_{i,j\in \mathbf{Y}_f,\; l_i=l_j} f(P_{i,j}, \mathbf{P})}{N_Y} \;\; \cdot \;\; \frac{\sum\limits_{i,j\in \mathbf{P}_f,\; l_i=l_j} f(P_{i,j}, \mathbf{Y})}{N_P}}
\end{split}
\end{equation}

\noindent
where $f(P_{i,j},\mathbf{A})$ is any function assessing the coherence between a path $P_{i,j}$ and a mask $\mathbf{A}$, returning 0 and 1 in the extreme cases of, respectively, no and full coherence. In our experiments, we consider two different possibilities for $f$. The first one, $f_H$, is based on the Hamming distance, $H$, a metric that counts the number of switches that are required to have two $n$-bit strings match. Notice that our problem can be interpreted as comparing two strings, since that: (i) a path $P_{i,j}$ of length $n$ can be seen as a $n$-length string of 1s; (ii) and the values that $\mathbf{A}$ takes at $\mathbf{x} \in P_{i,j}$ can also be represented as a $n$-length string, this time possibly containing both 0s and 1s, according to whether $\mathbf{A}$ takes the value of 0 or 1, respectively, at each of the path positions. Given that we want $f_H$ to produce values in the range $[0,1]$, and 1 to be equivalent to complete coherence, we define $f_H$ as: 

\begin{equation}
f_H(P_{i,j},\mathbf{A}) = \frac{n - H(P_{i,j}, \mathbf{A})}{n}
\end{equation}

\noindent
where $n >= 2$ is the number of pixels in the path $P_{i,j}$. An illustration demonstrating how $f_H$ is calculated is shown in Fig.~\ref{fig:hamming}.

\noindent
The second possibility is a binary function that only outputs 1 if the entire path is possible in the mask or, in other words, if the Hamming distance is 0:

\begin{equation}
f_F(P_{i,j},\mathbf{A}) =
\begin{cases}
1,                         & \text{if } H(P_{i,j}, \mathbf{A}) = 0  \\
0 & \text{otherwise}
\end{cases}
\end{equation}

\subsection{Practical considerations}

\begin{figure*}[t!]
	\begin{center}
		
		\begin{subfigure}{0.17\textwidth}
			\includegraphics[width=1\textwidth, height=1\textwidth]{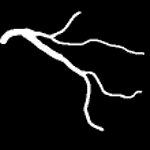}
			
			\medskip
			
			\includegraphics[width=1\textwidth, height=1\textwidth]{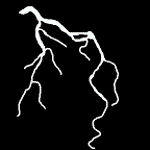}
			\caption{}
			\label{fig:aa}
		\end{subfigure}	
		\begin{subfigure}{0.17\textwidth}
			\includegraphics[width=1\textwidth, height=1\textwidth]{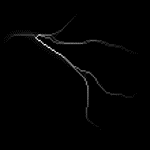}
			
			\medskip
			
			\includegraphics[width=1\textwidth, height=1\textwidth]{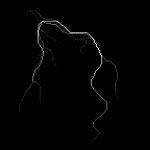}
			\caption{}
			\label{fig:bb}
		\end{subfigure}
		\begin{subfigure}{0.17\textwidth}
			\includegraphics[width=1\textwidth, height=1\textwidth]{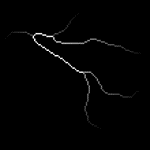}
			
			\medskip
			
			\includegraphics[width=1\textwidth, height=1\textwidth]{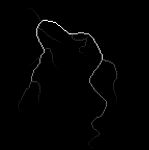}
			\caption{}
			\label{fig:cc}
		\end{subfigure}		
		\begin{subfigure}{0.17\textwidth}
			\includegraphics[width=1\textwidth, height=1\textwidth]{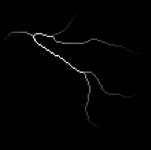}
			
			\medskip
			
			\includegraphics[width=1\textwidth, height=1\textwidth]{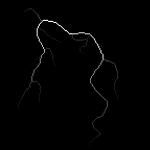}
			\caption{}
			\label{fig:dd}
		\end{subfigure}
	\end{center}
	\caption[The relative frequency each pixel is visited when considering all the possible paths to be taken in a tree.]{The relative frequency each pixel is visited when considering all the possible paths to be taken in a tree (a higher intensity is equivalent to a higher relative frequency). (a) Example coronary trees; (b) the frequency map when the minimum cost path between every two points of the tree is considered; (c) when the minimum cost path is retrieved from the centrelines for every two centreline points (strategy 1); and (d) when a distance function is used to promote the minimum cost path to follow the centrelines (strategy 2).}
	\label{fig:visited_pixels}
\end{figure*}

There are two important considerations to have into account regarding the similarity index defined before: i) to obtain $P_{i,j}$ we resort to a minimum cost path algorithm, which is likely to follow the boundaries of blood vessels when precaution is not taken (see Fig.~\ref{fig:bb}). This is an unwanted behaviour as it would be very sensitive to calibre-based errors and that is not the goal we seek with the proposed similarity index; ii) the number of possible paths grows exponentially with the number of pixels in $\mathbf{A}_f$, with a worst case complexity of $O(n^2/2)$. This turns the use of~\eqref{eq:metric_general3} impractical to evaluate real-world images which always have a significant number of blood vessel pixels.

Regarding the tendency of minimum cost path based approaches to follow the vessel boundaries, we discuss two possibilities to address this issue:

\begin{itemize}
	\item \textbf{Strategy 1}: instead of considering the complete set of vessel pixels $\mathbf{A}_f$, we can simply take into account the centreline pixels $\mathbf{A}_s$ and all the possible paths there. Concerning the third property we have defined earlier, this approach slightly reduces the relevance of larger vessels in comparison with the narrower ones. Even then, the mentioned natural properties of vascular trees (larger blood vessels ramify into smaller ones) promote centrelines of larger vessels to be visited more times (see Fig.~\ref{fig:cc});
	
	\item \textbf{Strategy 2}: another option is to use a non-linear distance function between the foreground (vessel) and background pixels, in order to promote the minimum cost path to follow the centrelines of the blood vessels. Comparing to the first strategy, this conserves slightly better the relevance of wider vessel segments (see Fig.~\ref{fig:dd}).
\end{itemize}

Fig.~\ref{fig:visited_pixels} shows two coronary trees and the number of times each pixel is visited when using the different strategies for employing the proposed similarity index. As illustrated, the two strategies proposed above effectively visit more frequently the main branches of the vascular trees, penalizing more any error in these segments.

Concerning the exponential complexity of the number of paths to be extracted, we can resort to a Monte Carlo approach to approximate~\eqref{eq:metric_general3}:

\begin{equation} \label{eq:approx_si}
	\begin{split}
		\tilde{m}&(\mathbf{P},\mathbf{Y})= \\
		& \bigg(\frac{1}{n} \sum\limits_{k=1}^n \; f(P_{i,j},\mathbf{P}),\; i,j \overset{i.i.d.}{\sim} U(\mathbf{Y}_f), \; l_i=l_j \;\; \cdot\\
		& \frac{1}{n} \sum\limits_{k=1}^n \; f(P_{i,j}, \mathbf{Y}),\; i,j \overset{i.i.d.}{\sim} U(\mathbf{P}_f), \; l_i=l_j \bigg)^{1/2}
	\end{split}
\end{equation}

\noindent
where $U(\mathbf{A}_f)$ denotes the Uniform distribution over a set of vessel pixels $\mathbf{A}_f$.

Regarding the example shown in Fig.~\ref{fig:cldice_invariance}, our proposed metric~\eqref{eq:metric_general3} evaluates $\mathbf{P}_1, \mathbf{P}_2$ and $\mathbf{P}_3$ with a score of, respectively, $0.984, 0.988,$ and $0.933$ (using Strategy 1). This is aligned with the properties we seek for a topological benchmark.

\section{A loss function for improving topological coherence}

In this section, we describe our proposed loss function which, besides focusing on the centrelines of the vascular tree, is also capable of distinguishing the errors mentioned in Property ii).

\subsection{Detecting errors that produce disjoint trees} \label{ssec:3_1}

We aim to design a loss function which, contrary to state-of-the-art losses, is able to highlight errors inducing disjoint trees, such that it can guide models towards producing blood vessel masks which are more topologically coherent. Errors leading to disjoint trees are nothing more than "holes" in a given blood vessel segment, therefore they may be filled by applying the morphological closing operator. 

\begin{figure}[t]
	\begin{center}
		
		\begin{subfigure}{0.27\columnwidth}
			\includegraphics[width=1\textwidth]{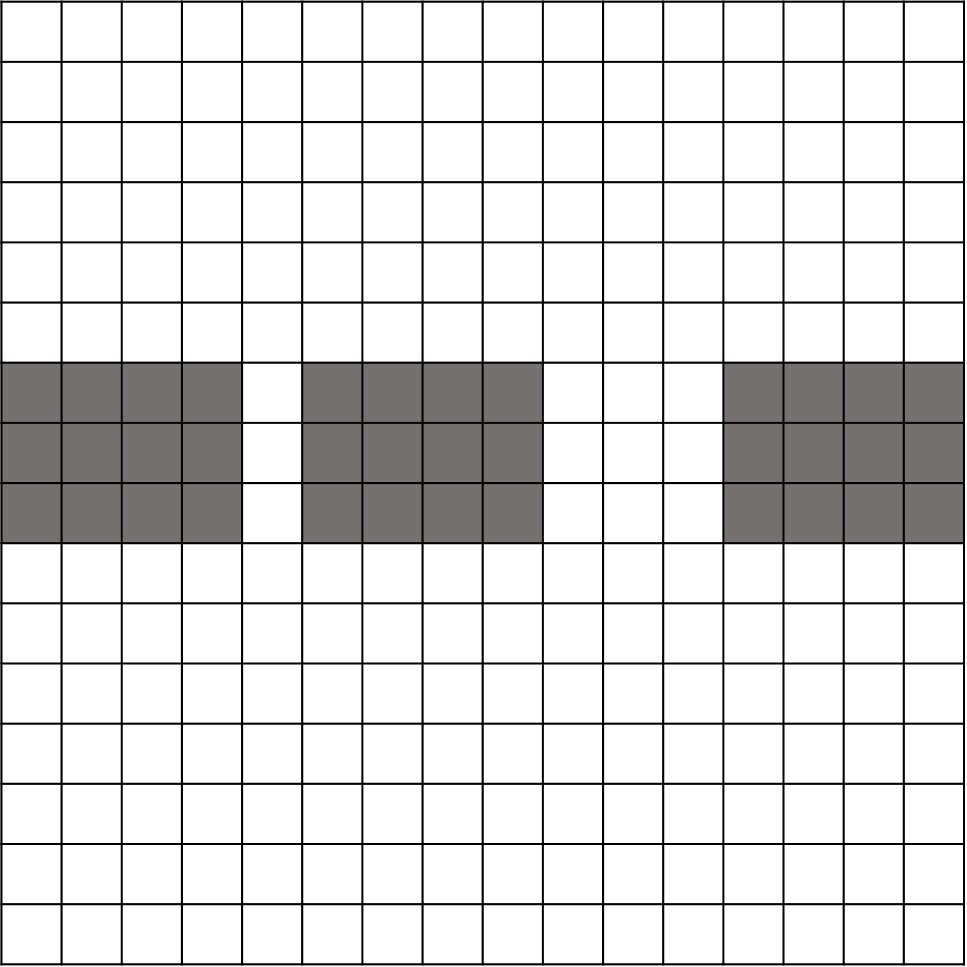}
			\caption{}
			\label{fig:ss_p}
		\end{subfigure}	\,
		\begin{subfigure}{0.27\columnwidth}
			\includegraphics[width=1\textwidth]{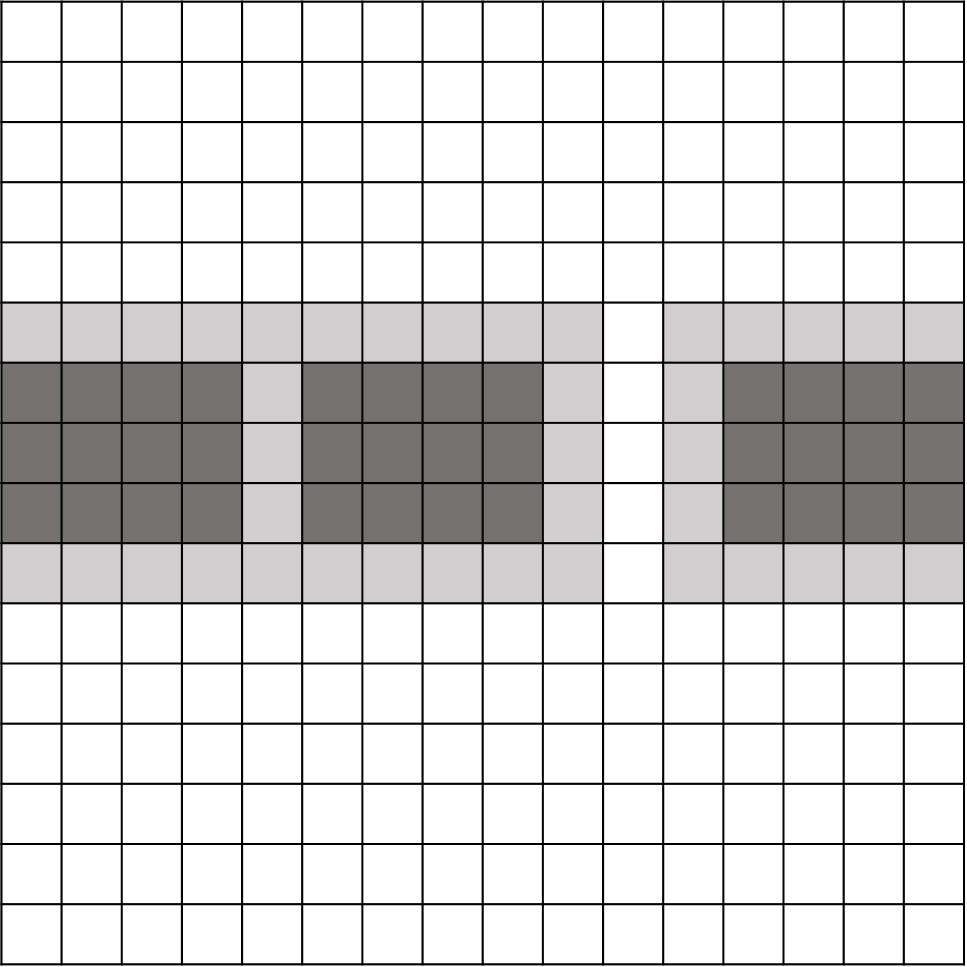}
			\caption{}
			\label{fig:ss_p_d1}
		\end{subfigure} \,
		\begin{subfigure}{0.27\columnwidth}
			\includegraphics[width=1\textwidth]{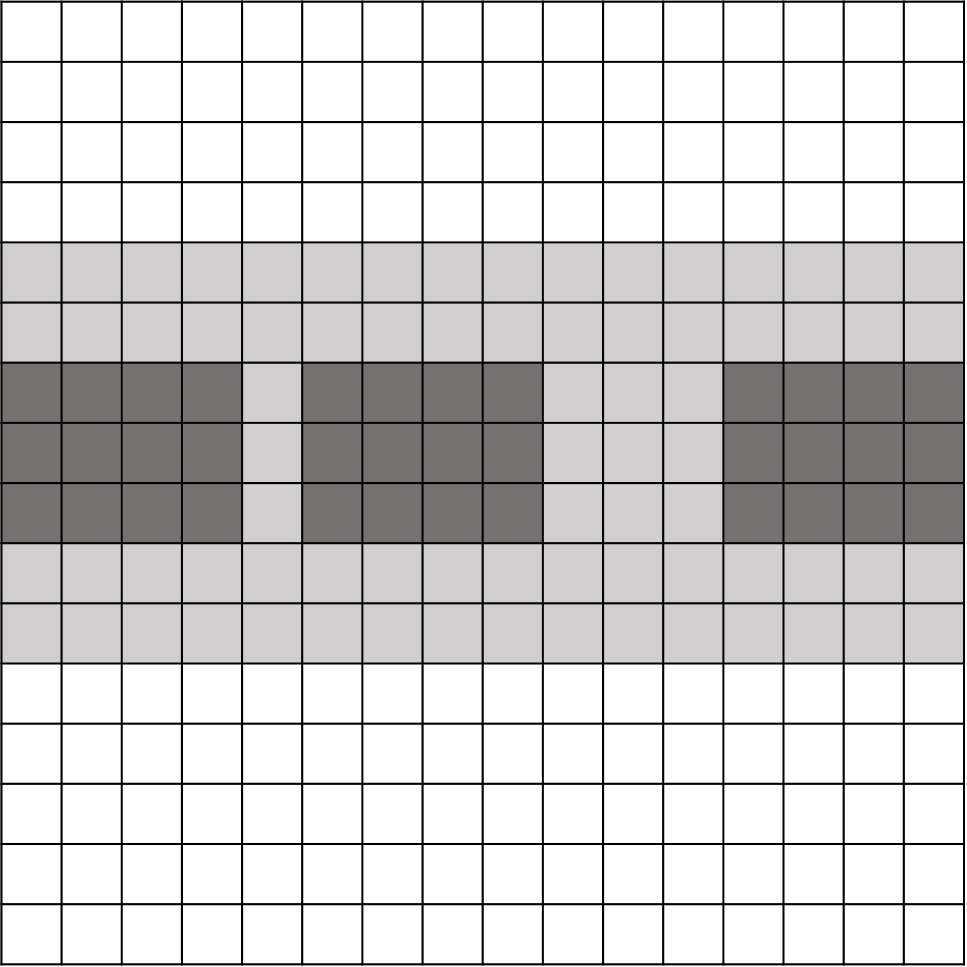}
			\caption{}
			\label{fig:ss_p_c1}
		\end{subfigure}
		
		\begin{subfigure}{0.27\columnwidth}
			\includegraphics[width=1\textwidth]{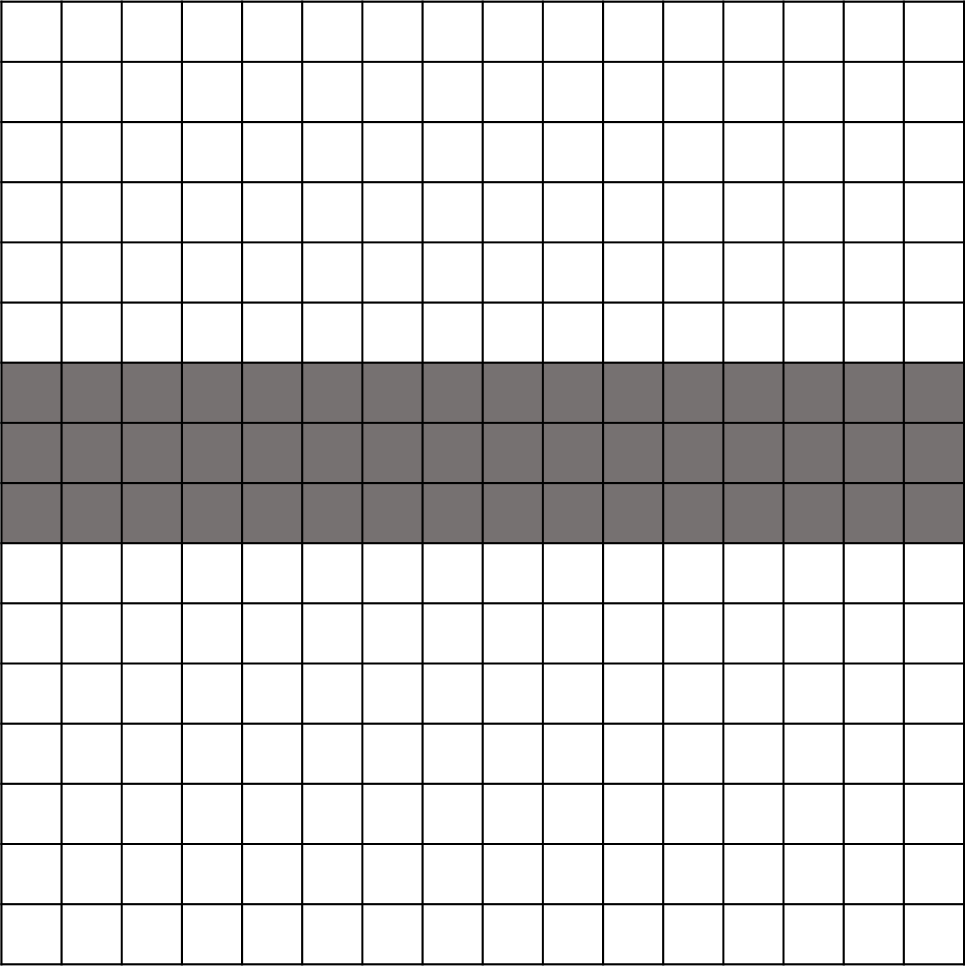}
			\caption{}
			\label{fig:ss_y}
		\end{subfigure} \,
		\begin{subfigure}{0.27\columnwidth}
			\includegraphics[width=1\textwidth]{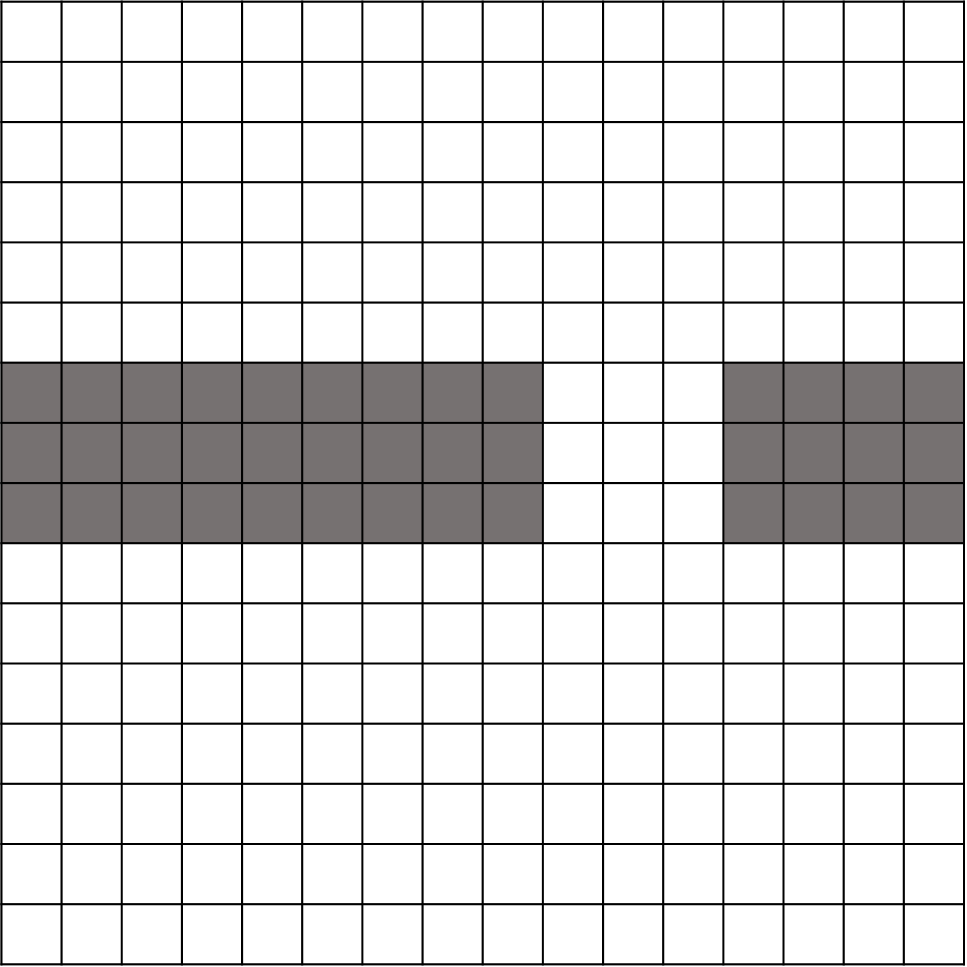}
			\caption{}
			\label{fig:ss_p_d2}
		\end{subfigure} \,
		\begin{subfigure}{0.27\columnwidth}
			\includegraphics[width=1\textwidth]{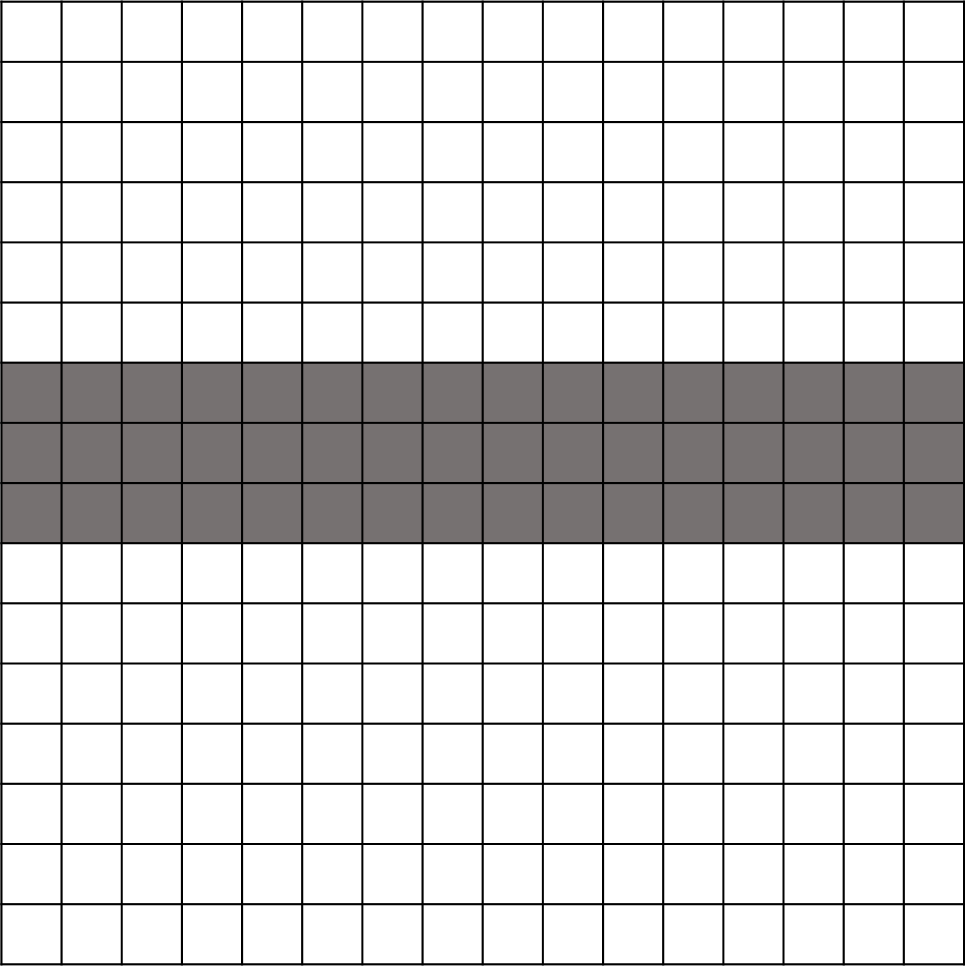}
			\caption{}
			\label{fig:ss_p_c2}
		\end{subfigure}
	\end{center}
	\caption[Connecting disjoint segments to recover the reference blood vessel segment, by means of a closing operation with a squared structuring element.]{Connecting disjoint segments to recover the reference blood vessel segment, by means of a closing operation with a squared structuring element. (a) Predicted, $\mathbf{P}$, and (d) reference, $\mathbf{Y}$, segmentations; morphological dilation of $\mathbf{P}$ with a SE of radius (b) 1, $\mathbf{P}_{D(1)}$, and (c) 2, $\mathbf{P}_{D(2)}$, where light grey represents the appended pixels; outputs of the respective closing operations, (e) $\mathbf{P}_{C(1)}$ and (f) $\mathbf{P}_{C(2)}$.}
	\label{fig:3_1}
\end{figure}

Let us start by considering the simple case of a single blood vessel segment being affected by this type of errors, as illustrated in Fig.~\ref{fig:3_1}.
The disjoint segments can be connected by employing a closing operator (dilation followed by erosion) using a structuring element (SE) of sufficient size. Let $\mathbf{A}_{D(r)}$ and $\mathbf{A}_{C(r)}$ be, respectively, the output of the morphological dilation and closing of $\mathbf{A}$ using a squared SE with radius $r$. In the example provided in Fig.~\ref{fig:3_1}, a radius of 2 would be necessary to connect all the disjoint segments.

\begin{figure}[t]
	\begin{center}		
		\begin{subfigure}{0.27\columnwidth}
			\includegraphics[width=1\textwidth]{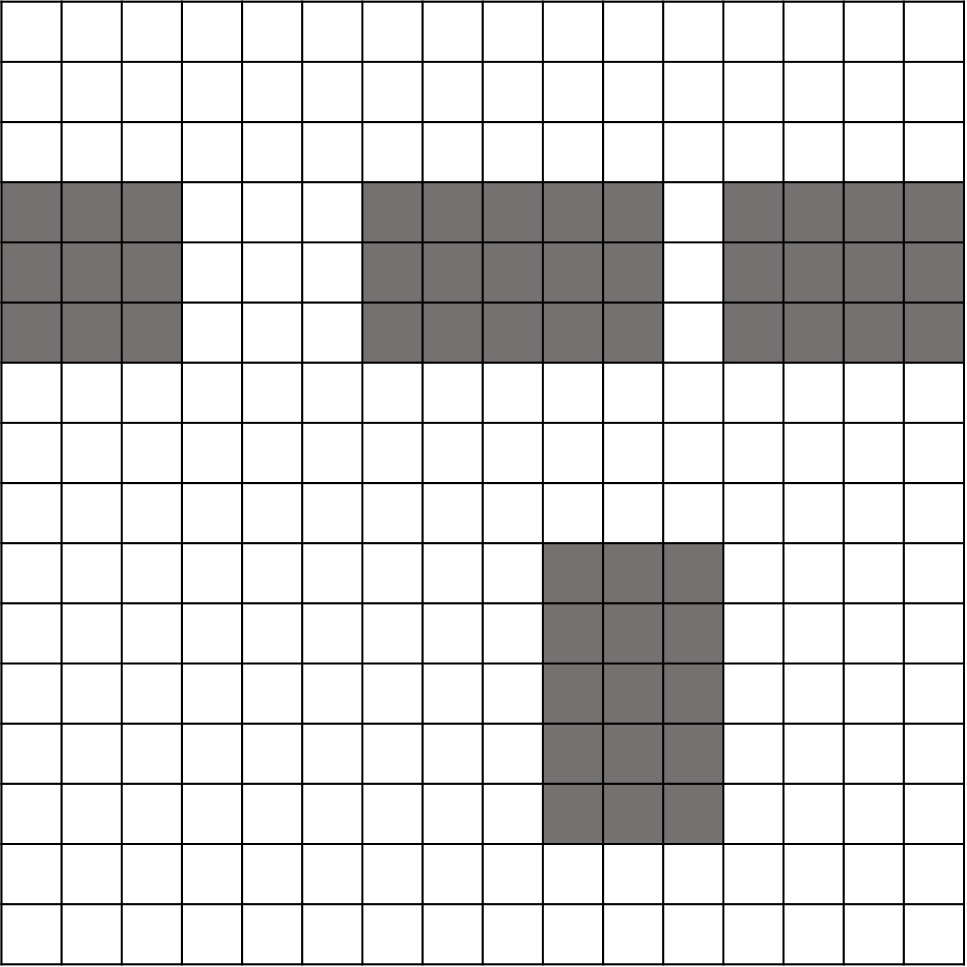}
			\caption{}
			\label{fig:ms_p}
		\end{subfigure} \,
		\begin{subfigure}{0.27\columnwidth}
			\includegraphics[width=1\textwidth]{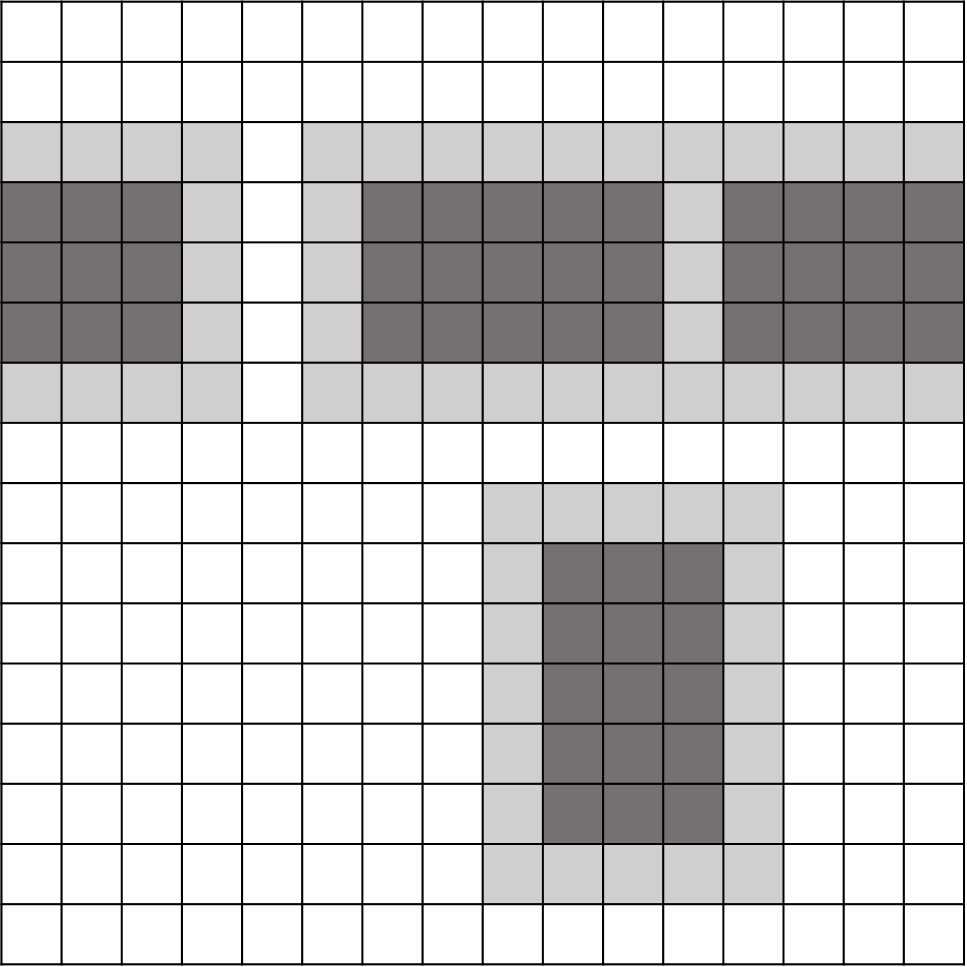}
			\caption{}
			\label{fig:ms_p_d1}
		\end{subfigure} \,
		\begin{subfigure}{0.27\columnwidth}
			\includegraphics[width=1\textwidth]{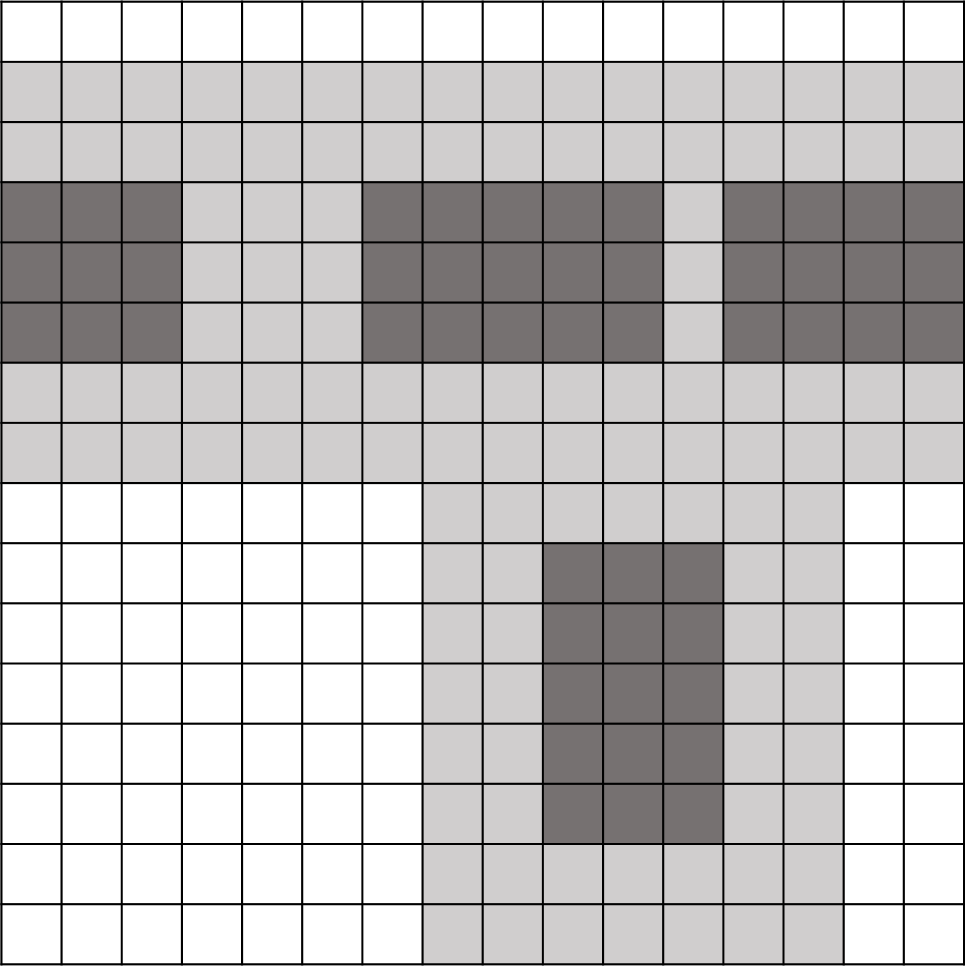}
			\caption{}
			\label{fig:ms_p_c1}
		\end{subfigure}
		
		\begin{subfigure}{0.27\columnwidth}
			\includegraphics[width=1\textwidth]{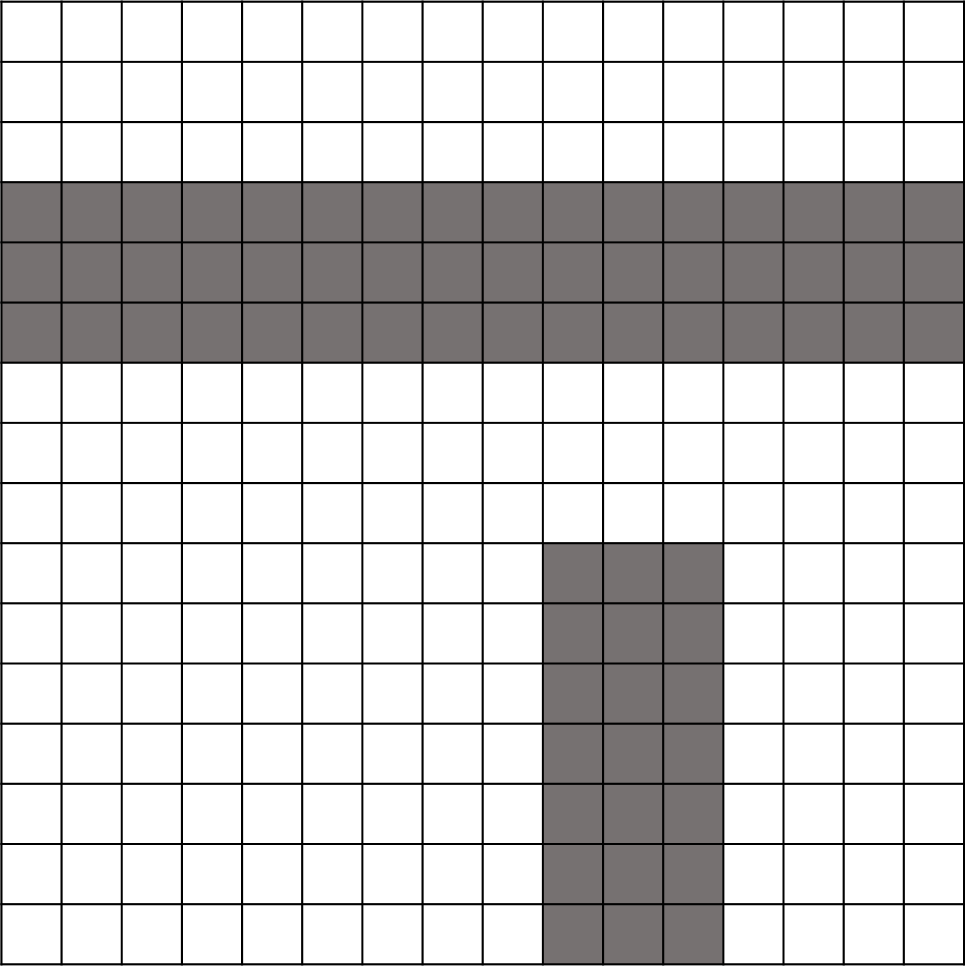}
			\caption{}
			\label{fig:ms_y}
		\end{subfigure} \,
		\begin{subfigure}{0.27\columnwidth}
			\includegraphics[width=1\textwidth]{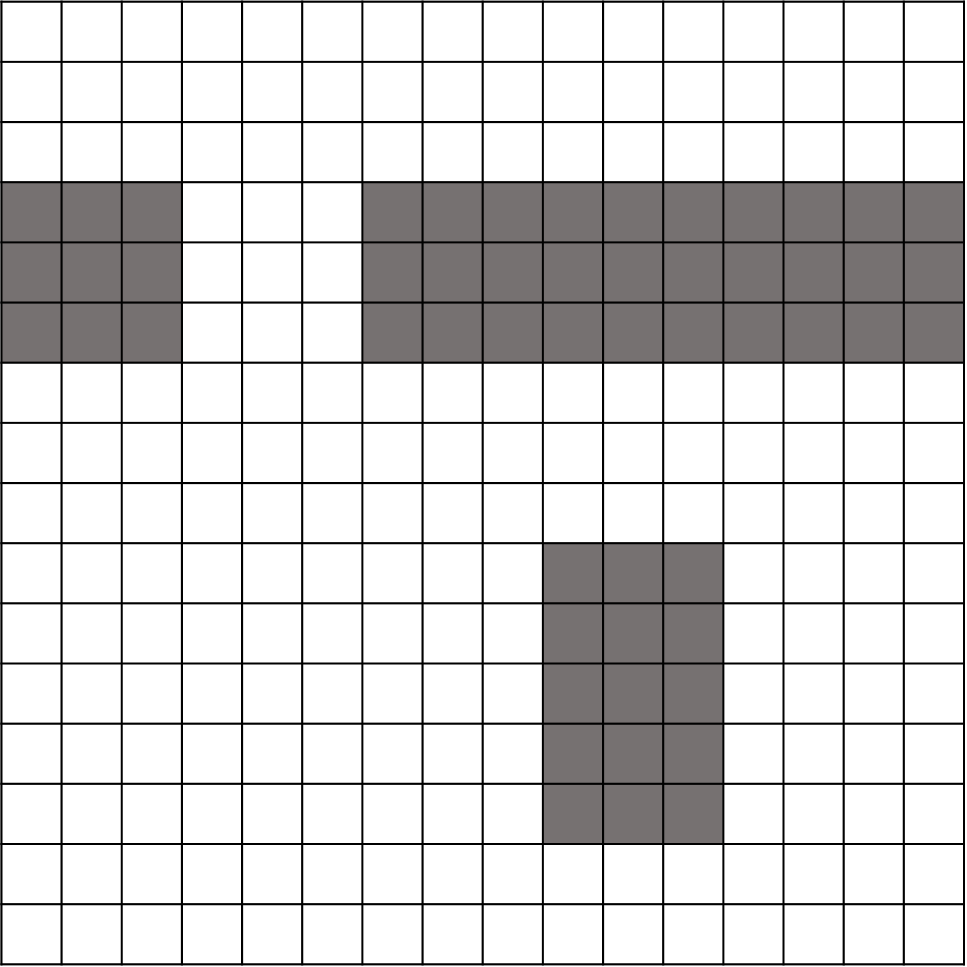}
			\caption{}
			\label{fig:ms_p_d2}
		\end{subfigure} \,
		\begin{subfigure}{0.27\columnwidth}
			\includegraphics[width=1\textwidth]{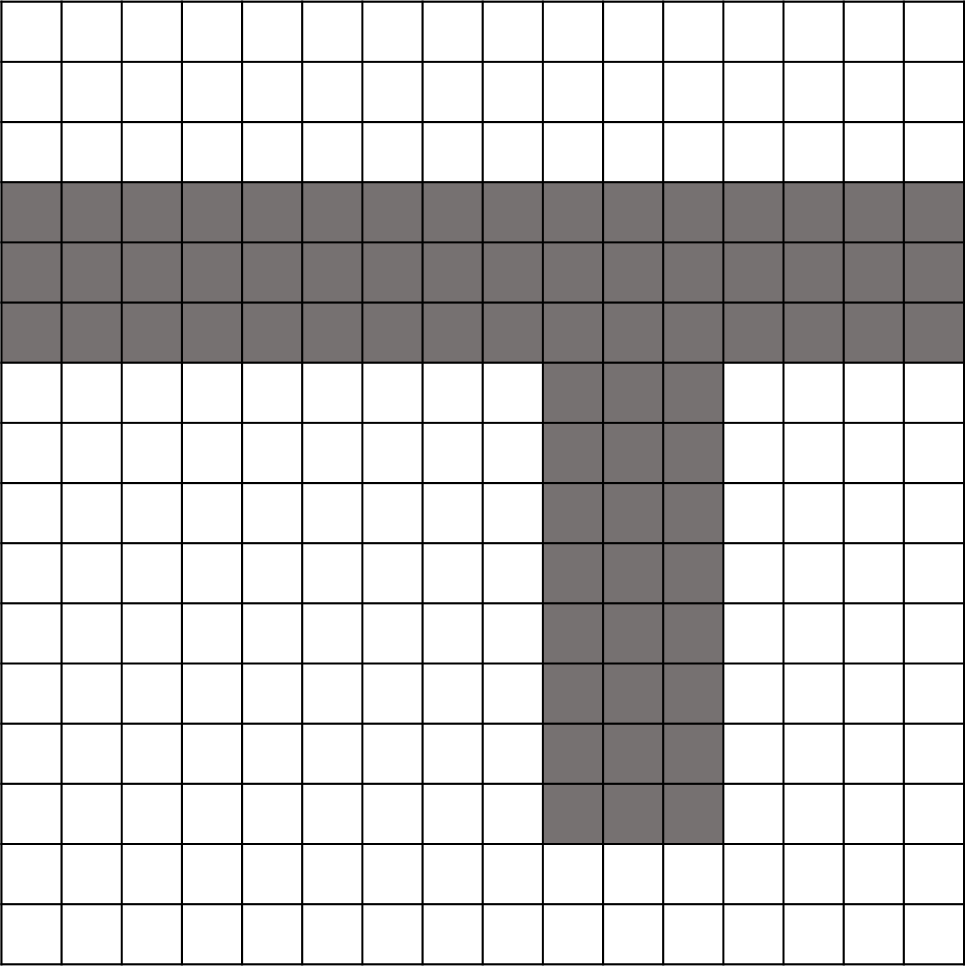}
			\caption{}
			\label{fig:ms_p_c2}
		\end{subfigure}		
	\end{center}
	\caption[Connecting disjoint segments to recover the reference blood vessel tree, showcasing the possibility of joining segments which are not connected in the reference segmentation.]{Connecting disjoint segments to recover the reference blood vessel tree, showcasing the possibility of joining segments which are not connected in the reference segmentation. (a) Predicted, $\mathbf{P}$, and (d) reference, $\mathbf{Y}$, segmentations; morphological dilation of $\mathbf{P}$ with SE of radius (b) 1, $\mathbf{P}_{D(1)}$, and (c) 2, $\mathbf{P}_{D(2)}$, where light grey represents the appended pixels; outputs of the respective closing operations, (e) $\mathbf{P}_{C(1)}$ and (f) $\mathbf{P}_{C(2)}$.}
	\label{fig:3_2}
\end{figure}

Most medical images of blood vessels (or even patches, small portions of these images) contain tree-like structures, not single vessel segments as illustrated in the previous simplistic scenario. Hence, the closing operation might merge disjoint segments which should not be connected at all (see Fig.~\ref{fig:3_2}), constituting an unwanted behaviour.
This effect is exacerbated when a large SE is required to connect segments far apart. Fortunately, as long as we have a reference segmentation, $\mathbf{Y}$, which is the case of supervised blood vessel segmentation, this limitation is easily circumvented. To do so, we pose the problem as joining separate segments only if the missing segment exists in the reference mask. In our experiments, we consider the centrelines $\mathbf{Y}_s$ instead, in order to put a larger focus on the graph structure of the vascular trees. Hence, we detect missing segments inducing disjoint trees as follows:

\begin{equation} \label{eq:3_1}
\mathbf{e}(\mathbf{P},\mathbf{Y}; r) = \big(\mathbf{P}_{C(r)} - \mathbf{P}\big)^2 \cdot \mathbf{Y_s}
\end{equation}

\begin{figure}[t]
	\begin{center}
		
		\begin{subfigure}{0.27\columnwidth}
			\includegraphics[width=1\textwidth]{ms_p.png}
			\caption{}
			\label{fig:ms_p_2}
		\end{subfigure} \,
		\begin{subfigure}{0.27\columnwidth}
			\includegraphics[width=1\textwidth]{ms_y.png}
			\caption{}
			\label{fig:ms_p_d1_}
		\end{subfigure} \,
		\begin{subfigure}{0.27\columnwidth}
			\includegraphics[width=1\textwidth]{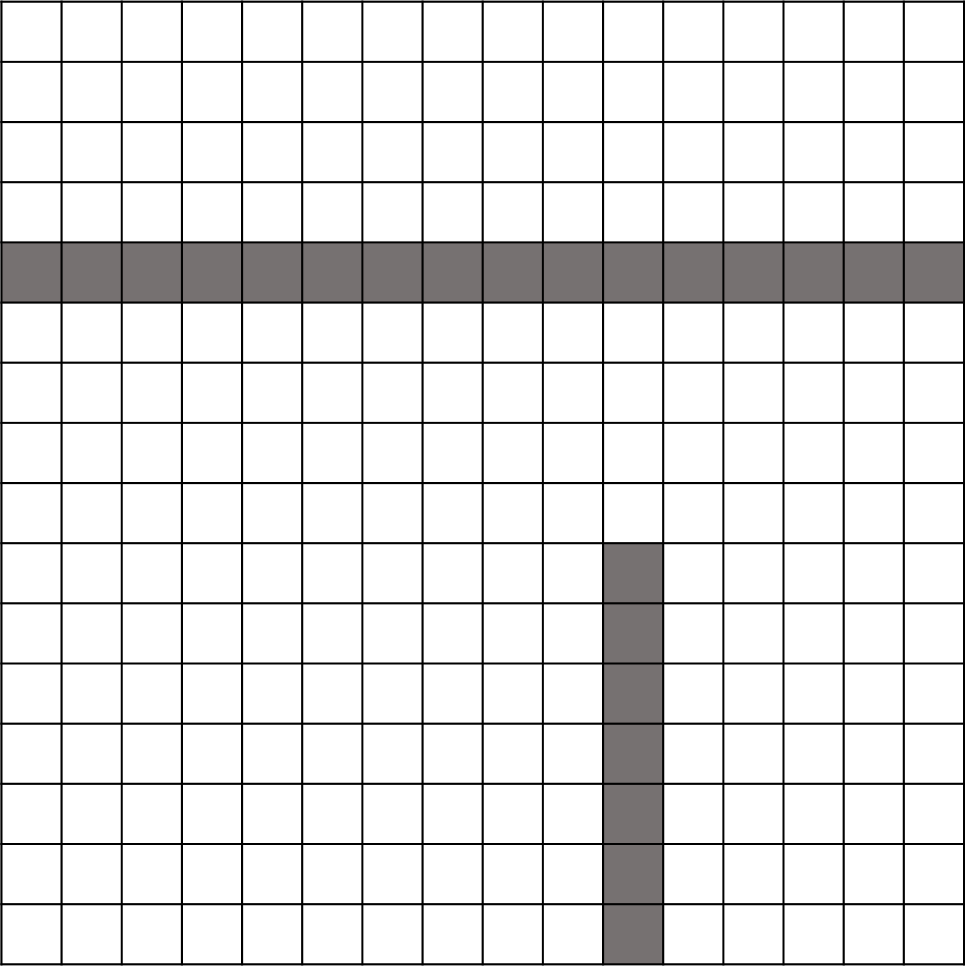}
			\caption{}
			\label{fig:ms_p_c1_}
		\end{subfigure}
		
		\begin{subfigure}{0.27\columnwidth}
			\includegraphics[width=1\textwidth]{ms_p_c2.png}
			\caption{}
			\label{fig:ms_y_2}
		\end{subfigure} \,
		\begin{subfigure}{0.27\columnwidth}
			\includegraphics[width=1\textwidth]{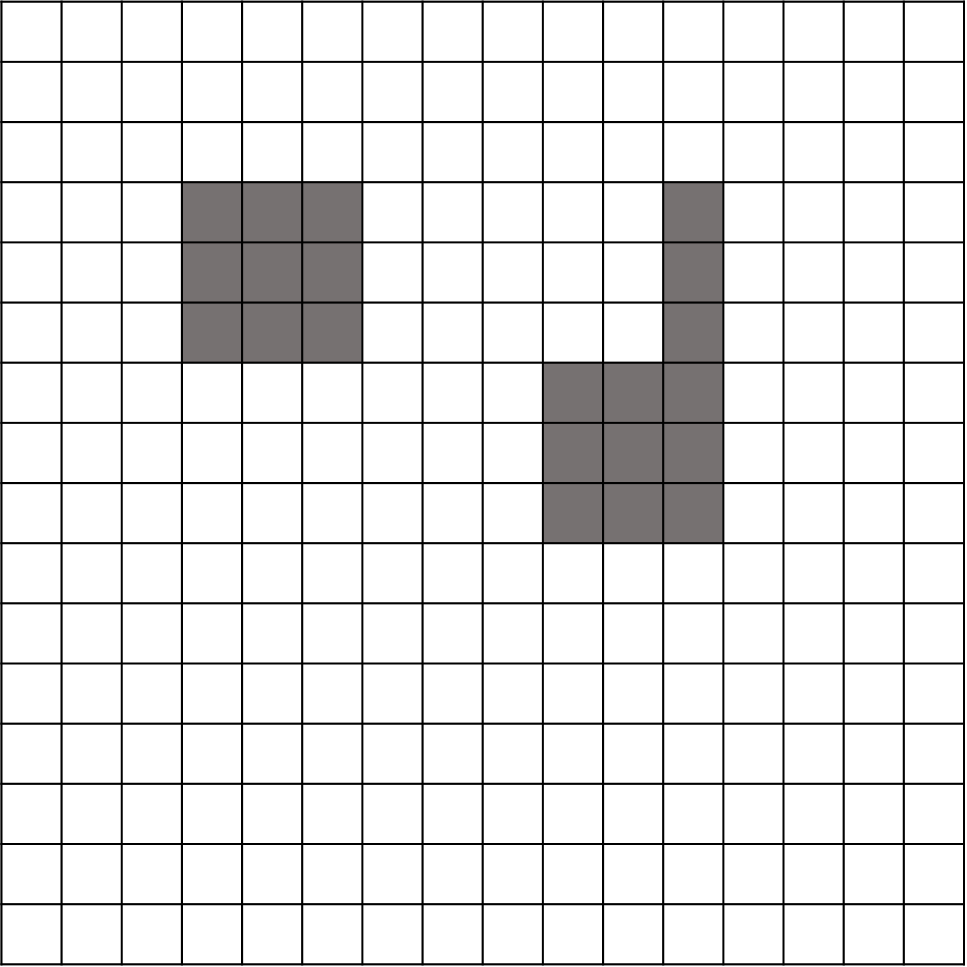}
			\caption{}
			\label{fig:ms_p_d2_2}
		\end{subfigure} \,
		\begin{subfigure}{0.27\columnwidth}
			\includegraphics[width=1\textwidth]{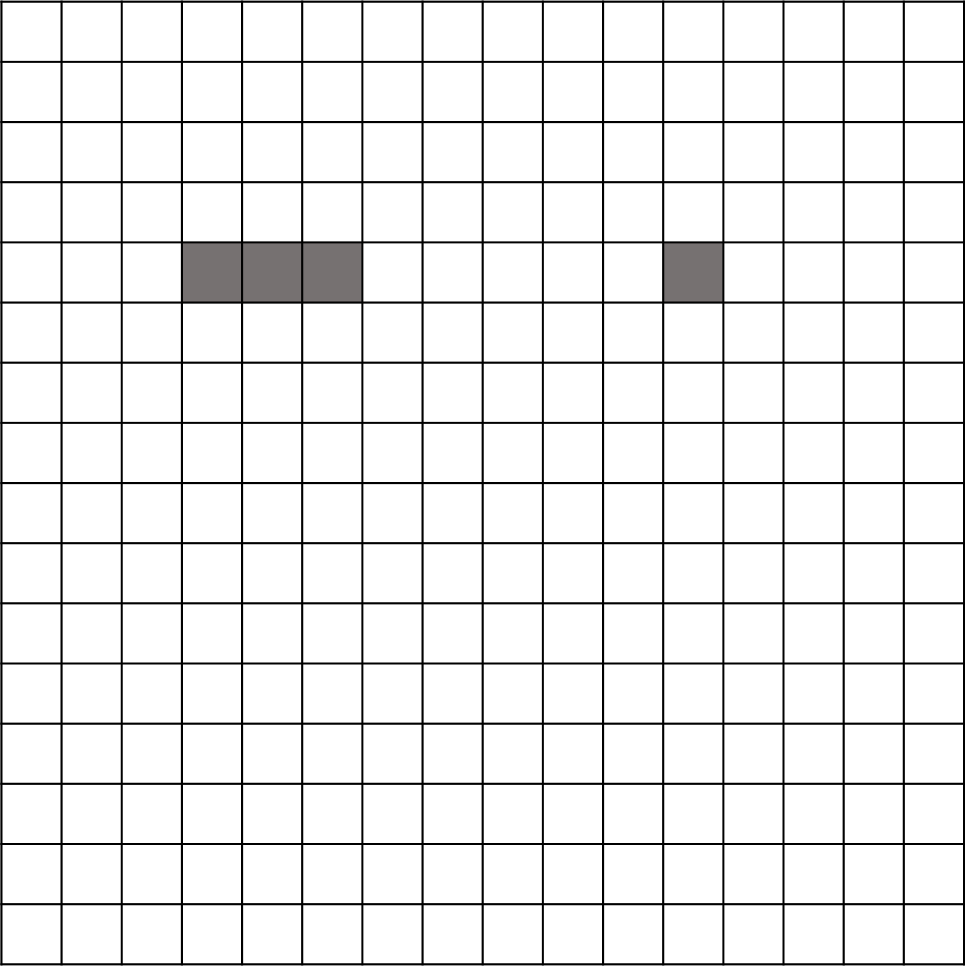}
			\caption{}
			\label{fig:ms_e2_2}
		\end{subfigure}
		
	\end{center}
	\caption[Detection of centreline errors inducing disjoint trees.]{Detection of centreline errors inducing disjoint trees. (a) Predicted, $\mathbf{P}$, and (b) reference, $\mathbf{Y}$, segmentations; (c) reference centrelines, $\mathbf{Y}_s$; (d) morphological closing of the prediction, $\mathbf{P}_{C(2)}$; (e) Squared difference between the latter and the original prediction, $(\mathbf{P}_{C(2)}-\mathbf{P})^2$; and (f) the centreline errors that produce disjoint trees, according to the reference mask, $\big(\mathbf{P}_{C(2)}-\mathbf{P}\big)^2\cdot \mathbf{Y}_s$.}
	\label{fig:3_3}
\end{figure}

\noindent
Fig.~\ref{fig:3_3} illustrates this approach on the example provided in Fig.~\ref{fig:3_2}.

\subsection{Weighting errors of different size} \label{ssec:3_2}

Despite identifying the errors that originate disjoint trees, according to~\eqref{eq:3_1}, their weight is being proportional to the length of the corresponding missing centreline. Therefore, following the example provided in Fig.~\ref{fig:3_3}, the larger missing segment would end having three times the weight of the smaller one. We argue that this might not be ideal for learning purposes. In our opinion, the model should learn that it is more likely that a small missing segment is in reality a false negative, than segments which are farther apart. Therefore, to reduce this weight bias towards larger missing segments, we consider weight normalization according to the length of the missing segment.

Let $S_r$ be a missing segment that can be filled through a closing operation considering a SE with radius $r$ or larger. The total error of this segment is given by:

\begin{equation}
e(S_r) = w_r \cdot l(S_r)
\end{equation}

\noindent
where $w_r$ is a scalar, and $l(S_r)$ is the length of the segment, being either $2r-1$ or $2r$. We seek a normalization such that the following inequality always holds:

\begin{equation} \label{eq:ineq_total_error}
w_r \cdot l(S_r) \geq w_{r+1} \cdot l(S_{r+1})
\end{equation}

\noindent
which states that a missing segment that can be filled with a SE with radius $r$ must have at least the same total error that a missing segment which can only be filled with a SE with radius $r+1$ or larger. By noticing again that the length of a missing segment $S_r$ can have two different values, the following equality is sufficient to hold inequality~\eqref{eq:ineq_total_error}:

\begin{equation} \label{eq:eq_total_error}
w_r\cdot (2r-1) = w_{r+1} \cdot (2 (r+1))
\end{equation}

Let us consider that the largest SE radius to be used is $r_{M}$ and that its associated weight $w_{r_M}$ is 1. It is now possible to consider an iterative approach that highlights all missing segments with length up to $2 r_M$ and normalizes their weights according to~\eqref{eq:eq_total_error}:

\begin{equation} \label{eq:normalized_error_eq}
\mathbf{e}(\mathbf{P},\mathbf{Y};r_M) = \mathlarger{\sum}_{r = r_M, r_{M-1}, \ldots, 1} \epsilon_r \cdot \big( \mathbf{P}_{C(r)} - \mathbf{P}\big)^2 \cdot \mathbf{Y_s}
\end{equation}

\noindent
with $\epsilon$ given as:

\begin{equation}
\epsilon_r=
\begin{cases}
w_{r_M},               & \text{if } r = r_M \\
w_r - w_{r+1}          & \text{otherwise}
\end{cases}
\end{equation}

\begin{figure*}[t]
	\begin{center}
		
		\begin{subfigure}{0.24\textwidth}
			\includegraphics[width=1\textwidth]{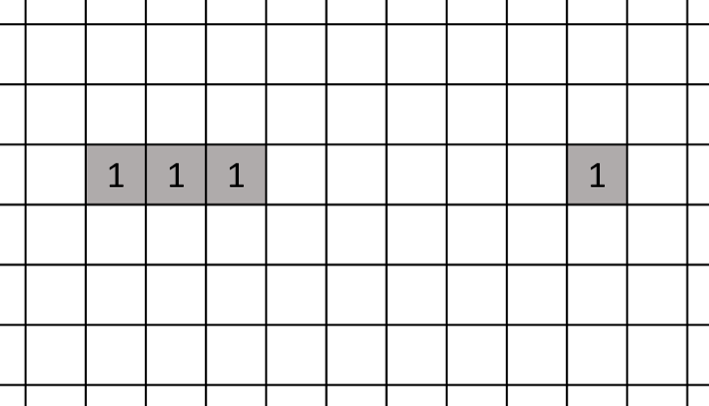}
			\caption{}
			\label{fig:err_norm_1}
		\end{subfigure}	\quad
		\begin{subfigure}{0.24\textwidth}
			\includegraphics[width=1\textwidth]{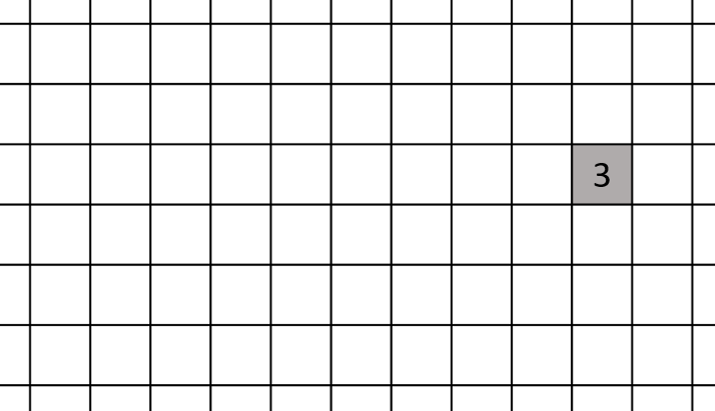}
			\caption{}
			\label{fig:err_norm_2}
		\end{subfigure} \quad	
		\begin{subfigure}{0.24\textwidth}
			\includegraphics[width=1\textwidth]{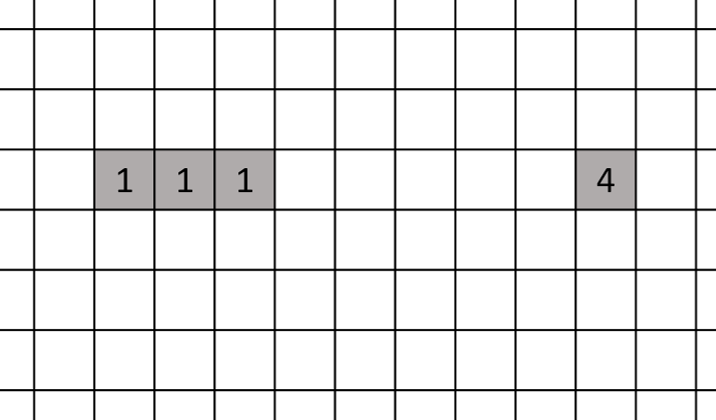}
			\caption{}
			\label{fig:err_norm_3}
		\end{subfigure}
	\end{center}
	\caption[Demonstration of the error normalization approach regarding the synthetic example considered in Fig. 5.9.]{Demonstration of the error normalization approach regarding the synthetic example considered in Fig.~\ref{fig:3_3}: (a) Missing segments detected when using a SE with radius $r_M$, which was set to 2 in this example, and the corresponding $\epsilon_2 = w_2 = 1$; (b) missing segments detected when using a SE with radius 1, with corresponding weight $w_1 = 4$, as given by~\eqref{eq:eq_total_error}, hence $\epsilon_1 = w_1 - w_2 = 3$; (c) total error according to~\eqref{eq:normalized_error_eq}.}
	\label{fig:example_error_norm}
\end{figure*}

Fig.~\ref{fig:example_error_norm} exemplifies how this normalizing approach would work for the synthetic example shown in Fig.~\ref{fig:3_3}, considering $r_M = 2$.

\subsection{Design of a topological loss} \label{ssec:3_3}

In subsections~\ref{ssec:3_1} and~\ref{ssec:3_2}, a method for highlighting errors inducing disjoint trees in a segmentation, according to a reference mask, was presented. Until now, false negative segments in the prediction segmentation have been focused. Nonetheless, it is also important to consider false positive detections joining segments which do not belong to the same tree in the reference mask. Note that this can be simply achieved via~\eqref{eq:normalized_error_eq} by switching the roles of the predicted segmentation and the reference mask. Therefore, we define two loss terms, one that seeks to minimize the amount of missing segments that induce disjoint trees, according to the reference mask:

\begin{equation}
L_{tsens}(\mathbf{P},\mathbf{Y};r_M)=\frac{\Big\| \mathlarger{\sum}_{r \in \mathbf{r}} \; \epsilon_r \cdot (\mathbf{P}_{C(r)} - \mathbf{P}) \cdot \mathbf{Y}_s \Big\|_1}{\| \mathbf{Y}_s \|_1}
\end{equation}

\noindent
and a second one, promoting that no false connections are introduced, again having into account the reference mask:
\begin{equation}
L_{tprec}(\mathbf{P},\mathbf{Y};r_M)=\frac{\Big\| \mathlarger{\sum}_{r \in \mathbf{r}} \; \epsilon_r \cdot (\mathbf{Y}_{C(r)} - \mathbf{Y}) \cdot \mathbf{P}_s \Big\|_1}{\| \mathbf{P}_s \|_1}
\end{equation}

\noindent
where the denominators are introduced for normalizing purposes only and not considered during the loss gradient calculation and, for simplicity of notation, $\mathbf{r} = r_M, r_{M-1}, \ldots, 1$. Both terms are combined into our proposed topological loss:

\begin{equation} \label{eq:proposed_topo_loss}
\begin{split}
L_{topo}(\mathbf{P},\mathbf{Y};\alpha, r_M) = & \, \alpha \cdot L_{tsens}(\mathbf{P},\mathbf{Y};r_M) + \\
& (1-\alpha) \cdot L_{tprec}(\mathbf{P},\mathbf{Y};r_M)
\end{split}
\end{equation}

\noindent
where $\alpha \in [0, 1]$ determines the relative importance of each term.

In order to easily integrate the morphological operators into the model training procedure, we consider the neural dilation and erosion layers introduced in~\cite{a36}. Note that, instead of having to design multiple neural layers to perform morphological closings with SE of different radius, we can simply cascade neural layers implementing morphological operations with a SE of radius 1. This is possible since the following holds:

\begin{equation}
\mathbf{A}_{D(r)} = \mathbf{A}_{D(1)}^{r}
\end{equation}

\begin{equation}
\mathbf{A}_{C(r)} = \big({\mathbf{A}_{D(r)}}\big)_{E(1)}^{r}
\end{equation}

\noindent
where $\mathbf{A}_{M(1)}^r$ denotes $r$-consecutive uses of a morphological operator $M$ with a SE with radius 1 in segmentation mask $A$.
\section{Experiments}~\label{sec:experiments}

To assess the contribute of the proposed loss, $L_{topo}$~\eqref{eq:proposed_topo_loss}, in the scenario of blood vessel segmentation, we analysed how the performance of an U-net model~\cite{a37} varies according to the loss function that is minimized during training.

The first two baseline loss functions considered are the typically used soft-Dice loss, $L_{dice}$, and the BCE loss, $L_{bce}$:

\begin{equation}
L_{dice}(\mathbf{P},\mathbf{Y}) = 1 - 2 \cdot \frac{ \mathlarger{\sum}_{i} \; \mathbf{P}_i \cdot \mathbf{Y}_i}{\mathlarger{\sum}_i \; \mathbf{P}_i^2 + \mathbf{Y}_i^2}
\end{equation}

\begin{equation}
	\begin{split}
		L_{bce}(\mathbf{P},\mathbf{Y};\alpha) = & - \frac{1}{N} \mathlarger{\mathlarger{\sum}}_{i=1}^N \Big( \alpha \cdot \mathbf{Y}_i \cdot \log(\mathbf{P}_i) + \\		
		& (1-\alpha) \cdot (1-\mathbf{Y}_i) \cdot \log (1-\mathbf{P}_i) \Big)
	\end{split}
\end{equation}

\noindent
where $N$ is the total number of pixels in the image, and $\alpha \in [0, 1]$ determines the relative weight of each class (blood vessel and background).

Given the results achieved in~\cite{a32}, the clDice loss, a term that emphasizes the consistency along the centrelines of the blood vessels, was shown to promote the model to learn masks which are more similar to the reference one topological-wise. Therefore, we also use the loss proposed in~\cite{a32} as a baseline:

\begin{equation} \label{eq:cldice_loss}
	\begin{split}
		L_{cldice}(\mathbf{P},\mathbf{Y};\alpha) = & \alpha \cdot L_{dice}(\mathbf{P},\mathbf{Y}) + \\
		& (1-\alpha) \cdot (1 - clDice(\mathbf{P},\mathbf{Y}))
	\end{split}
\end{equation}

\noindent
where $clDice(\mathbf{P},\mathbf{Y})$ is the criterion defined in~\cite{a32}, and $\alpha$ determines the relative weight given to each loss component. A loss mimicking the ideas behind $L_{cldice}$, but using the BCE criterion instead, was also regarded in the experiments:

\begin{equation} \label{eq:clbce_loss}
	\begin{split}
		& L_{clbce} (\mathbf{P},\mathbf{Y};\alpha;\beta) = \\
		& -\frac{1}{N} \mathlarger{\mathlarger{\sum}}_{i=1}^{N} \Big( (a \cdot \mathbf{Y}_i + \beta \cdot \mathbf{Y}_{s_i}) \cdot \log(\mathbf{P}_i) + \\
		& \big((1-\alpha) \cdot (1-\mathbf{Y}_i) + (1-\beta) \cdot \mathbf{P}_{s_i}\big) \cdot \log (1-\mathbf{P}_i) \Big)
	\end{split}
\end{equation}

\noindent
with $\beta$ controlling the compromise between false positive and false negative centreline detections.

Having described the losses that will be used as benchmark in the experiments, the proposed loss function is now detailed. Since we recognize the relevance of emphasizing the centrelines of the vessels, we let our proposed loss build upon $L_{cldice}$ \eqref{eq:cldice_loss} and $L_{clbce}$ \eqref{eq:clbce_loss}, and extend them by including an additional term pertaining to the loss proposed in subsection~\ref{ssec:3_3}:

\begin{equation}
	\begin{split}
		L_{propdice}(\mathbf{P},\mathbf{Y};\alpha_1,\alpha_2,c,r) =& L_{cldice}(\mathbf{P},\mathbf{Y};\alpha_1) +\\
		& c \cdot L_{topo}(\mathbf{P},\mathbf{Y};\alpha_2,r)
	\end{split}
\end{equation}

\begin{equation}
	\begin{split}
		L_{propbce}(\mathbf{P},\mathbf{Y};\alpha_1,\alpha_2,\beta,c,r) =& L_{clbce}(\mathbf{P},\mathbf{Y};\alpha_1, \beta) +\\
		& c \cdot L_{topo}(\mathbf{P},\mathbf{Y};\alpha_2,r)
	\end{split}
\end{equation}

\noindent
where $c$ sets the relevance of the proposed topological term in the overall loss, $\alpha_2$ controls the compromise between $L_{tsens}$ and $L_{tprec}$, and $r$ defines the maximum radius to be considered in the involved closing operations.

\subsection{Datasets and Model Evaluation} \label{data_metrics}

Regarding the data used in the experiments, in addition to the retinal vessel segmentation benchmarks commonly used (DRIVE~\cite{a21}, STARE~\cite{a33}, and CHASE~\cite{a34}), we have also considered a dataset containing coronary angiograms~\cite{a35}.

Concerning metrics and similarity indices to assess the performance of the different models, we consider not only the typically used metrics (AUC, accuracy, sensitivity, and specificity), but also the clDice score~\cite{a32}, and both variants of the proposed approximate topological similarity index~\eqref{eq:approx_si}, $\tilde{m}_H$ and $\tilde{m}_F$. The Monte Carlo approximation was performed with $n=1000$.

\subsection{Implementation Details}

From the 20 images comprising the training set of DRIVE, 4 were set aside for validation purposes. The original test set remained unchanged and was used for that stage. Regarding STARE, from a total of 20 images, 3 and 5 were reserved for, respectively, validation and testing phases. Concerning CHASE, from the available 28 images, 4 and 10 were used for the validation and test steps. Finally, for the CORONARY dataset, which contains 134 coronary angiograms, we considered 20 and 30 images for validating and testing, respectively, the trained models. Throughout training, 128$\times$128 patches were fed to the model, whereas, during validation and testing, the entire images were given. Data augmentation during training comprised flipping and rotation transformations, and the addition of an intensity bias.

The U-net model~\cite{a37} was implemented according to its original description and 100 batches containing 8 patches each were fed every epoch. The Adam optimizer was used to update the model parameters, with an initial learning rate of $10^{-4}$. The loss in the validation set was measured every 25 epochs. Every 50 epochs the learning rate was decreased by a factor of 0.1, being the training process terminated after 200 epochs. We kept the set of parameters leading to a minimum loss over the evaluations performed in the validation set. To regularize the weights of the model, excluding the bias of neurons, we considered $L_2$-regularization with a coefficient of $10^{-5}$.

Two different configurations of $L_{bce}$ were considered, one where an equal weight was given to both classes ($\alpha=0.5$), $L_{bceu}$, and another one weighting significantly more errors in blood vessel pixels ($\alpha=0.7$), $L_{bcew}$. This weight has been found appropriate in previous experiments. $L_{cldice}$ was configured following~\cite{a32}, by setting $\alpha=0.5$. In our proposed variant, $L_{propdice}$, we also consider $\alpha_1=0.5$, then we set $r=10$, $c=0.1$, and test different values for $\alpha_2$. The centreline aware BCE variants, $L_{clbceu}$, and $L_{clbcew}$ were parametrised with $\beta=0.5$, and $\alpha$ of, respectively, 0.5 and 0.7. Our proposed extension, $L_{propbce}$, considers $\alpha_1=0.7, \beta=0.5, r=10, c=0.1$ and, again, several values are tested for $\alpha_2$.

To disregard randomness involved in training using a GPU and other processes such as batch generation, we have performed deterministic training and evaluation. To further highlight the differences between the losses, we have run 2 times each of them, by picking 2 different seeds, guaranteeing that among experiments with the same seed, everything was constant except the loss function being optimized. An NVIDIA GeForce RTX 2080 Ti GPU was used to conduct the experiments. Our implementation in PyTorch of the model, training procedure, and described losses, will be made available upon acceptance of this manuscript.

\section{Results and Discussion}

The average value of the performance obtained when minimizing the different losses, concerning the benchmarks specified in subsection~\ref{data_metrics}, is shown in Table~\ref{tab_results_topo_loss}.

\begin{table*}[t]
	\begin{center}
	\captionof{table}[Effect of the proposed topological loss term in the performance of the models.]{Performance of the models, in percentage, averaged over 2 runs. AUC and \textit{acc}, stand for, respectively, area under the roc curve, and accuracy. clDice is the metric proposed in~\cite{a32}, and $\tilde{m}_H$ and $\tilde{m}_F$ are the two approximate topological similarity indices presented in section~\ref{sec:2}. The larger all of these indicators, the better the performance of the model. The best obtained performance for each indicator, model family, and considered database is highlighted in bold. The best model for a given database and indicator is underlined.}
	\begin{tabular}{@{}cc>{\centering\arraybackslash}p{11mm}>{\centering\arraybackslash}p{11mm}>{\centering\arraybackslash}p{11mm}>{\centering\arraybackslash}p{11mm}>{\centering\arraybackslash}p{11mm}>{\centering\arraybackslash}p{11mm}>{\centering\arraybackslash}p{11mm}>{\centering\arraybackslash}p{11mm}>{\centering\arraybackslash}p{11mm}@{}}
		\toprule
		&        & $L_{dice}$ & $L_{cldice}$ & $L_{propdice}$ & $L_{bceu}$ & $L_{clbceu}$ & $L_{propbceu}$ & $L_{bcew}$ & $L_{clbcew}$ & $L_{propbcew}$ \\ \midrule
		\multirow{5}{*}{DRIVE} & AUC & \textbf{97.5} & 95.8 & 96.0 & \underline{\textbf{97.8}} & 97.6 & 97.6 & \textbf{97.6} & 97.4 & 97.4 \\
		& acc    & \textbf{95.2}  & 94.6 & 93.5 & \underline{\textbf{95.5}} & 95.2 & 94.0 & \textbf{94.8} & 94.4 & 93.6 \\
		& clDice & 82.2 & 84.4 & \textbf{85.2} & 81.4 & 81.6 & \textbf{84.6} & 82.6 & 83.0 & \textbf{83.4} \\
		& $\tilde{m}_H$ & 93.0 & 94.2 & \underline{\textbf{94.8}} & 93.0 & 92.8 & \textbf{94.3} & 93.0 & 92.8 & \textbf{94.6}  \\
		& $\tilde{m}_F$ & 18.8 & 19.3 & \underline{\textbf{21.0}} & 18.9 & 17.0 & \textbf{19.4} & 18.3 & 17.6 & \textbf{20.4}  \\ \midrule
		\multirow{5}{*}{STARE} & AUC & \textbf{98.6} & 96.6 & 96.8 & \textbf{98.6} & \textbf{98.6} & \textbf{98.6} & \underline{\textbf{98.8}} & \underline{\textbf{98.8}} & 98.7 \\
		& acc    & \underline{\textbf{97.0}} & 96.8 & 96.6 & \underline{\textbf{97.0}} & \underline{\textbf{97.0}} & 96.4 & \textbf{96.6} & 96.5 & 95.8 \\
		& clDice & 87.3 & 87.7 & \underline{\textbf{88.6}} & 86.7 & 87.5 & \textbf{88.4} & 87.6 & 87.7 & \textbf{88.4} \\
		& $\tilde{m}_H$ & 94.6 & 94.4 & \textbf{95.2} & 94.4 & 94.7 & \underline{\textbf{95.7}} & 94.6 & 95.2 & \textbf{95.6} \\
		& $\tilde{m}_F$ & 29.4 & 30.6 & \textbf{31.2} & 30.8 & 30.3 & \textbf{34.6} & 31.0 & 34.0 & \underline{\textbf{35.6}} \\ \midrule
		\multirow{5}{*}{CHASEDB1} & AUC & \textbf{97.2} & 94.4 & 94.8 & 97.8 & \underline{\textbf{98.0}} & 97.8 & \textbf{97.3} & 96.8 & 96.8 \\
		& acc & \textbf{95.5} & 95.4 & 94.6 & \underline{\textbf{96.2}} & \underline{\textbf{96.2}} & 95.8 & \textbf{94.8} & 94.1 & 93.6 \\
		& clDice & 78.0 & 80.0 & \textbf{80.6} & 80.8 & \underline{\textbf{81.9}} & 80.8 & \textbf{77.2} & 73.8 & 76.3 \\
		& $\tilde{m}_H$ & 85.6 & 87.2 & \textbf{87.7} & 89.6 & 89.4 & \underline{\textbf{90.1}} & 85.0 & 82.4 & \textbf{85.9} \\
		& $\tilde{m}_F$ & 12.5 & 12.9 & \textbf{14.5} & 15.9 & 15.0 & \underline{\textbf{17.0}} & 11.6 & 9.2 & \textbf{14.2} \\ \midrule
		\multirow{5}{*}{CORONARY} & AUC & \textbf{98.4} & 96.9 & 96.8 & 98.8 & \textbf{98.9} & 98.8 & 99.0 & \underline{\textbf{99.1}} & 98.9 \\
		& acc    & \textbf{97.6} & 97.4 & 97.0 & \underline{\textbf{97.7}} & \underline{\textbf{97.7}} & 96.8 & 97.1 & \textbf{97.2} & 96.4 \\
		& clDice & 84.6 & \underline{\textbf{85.6}} & 85.4 & 84.0 & \textbf{84.2} & 84.1 & 83.4 & 84.2 & \textbf{84.8} \\
		& $\tilde{m}_H$ & 89.3 & 90.1 & \textbf{90.6} & 89.4 & 89.8 & \textbf{90.4} & 89.4 & 90.4 & \underline{\textbf{91.1}} \\
		& $\tilde{m}_F$ & 39.3 & 42.0 & \underline{\textbf{43.2}} & 39.2 & 40.4 & \textbf{42.8} & 41.0 & \textbf{42.6} & 41.9 \\ \bottomrule		
	\end{tabular}
	\label{tab_results_topo_loss}
	\end{center}
\end{table*}

For each loss that includes the proposed cost function and for each dataset, we show only the results for the parametrisation achieving larger average value of $\tilde{m}_H$ and $\tilde{m}_F$. The best parametrisations are highlighted with a filled circle in the graphics concerning the ablation studies, which can be found at \url{www.github.com/rjtaraujo/topo-bloodvessel}. The ideal value of $\alpha_2$ varies according with the dataset and the loss that is being considered. The larger the $\alpha_2$ value, the larger the focus on errors inducing disjoint trees over those merging separate trees. Even though a pattern is not clear for $L_{propdice}$, $L_{propbceu}$ and $L_{propbcew}$ seem to benefit from, respectively, larger and smaller $\alpha_2$ values. The weighted BCE loss already weights significantly blood vessel locations, such that it seems plausible that a big focus on errors inducing disjoint trees would hurt the equilibrium between these errors and those merging distinct trees. Following the same rationale, models based on the unweighted BCE typically have difficulty in effectively capturing some of the vessel branches; therefore, it is likely that they benefit more from the use of large $\alpha_2$ values. Despite this variability regarding the optimal $\alpha_2$, the ablation studies show that, for a particular family of models, it is possible to find $\alpha$ values which lead to improvements in all the datasets when compared with the centreline-aware baselines. By analysing the graphics, it is also possible to conclude that there is not a direct relation between $\tilde{m}_H$ and $\tilde{m}_F$, such that it is possible to have models which improve one of them while having their performance decreased in the other one. This shows that, even though our proposed general similarity index~\eqref{eq:metric_general3} has fixed properties related to the graph of the vascular trees, the considered function $f(P_{i,j},\mathbf{A})$ also plays a role on the properties that are highlighted during the benchmark.

Analysing the patterns found for each of the loss function families, and starting with the Dice one, $L_{dice}, L_{cldice},$ and $L_{propdice}$, the impact of the $soft-clDice$ criterion in the AUC is evident, since the models that include the minimization of this criterion have significantly decreased AUC in comparison to $L_{dice}$. The accuracy was also slightly higher in the $L_{dice}$ experiments; however, we believe this is not a crucial metric for blood vessel segmentation, as class imbalance exists, and models achieving better compromises between sensitivity and specificity tend to have lower accuracies. This trend is observable in the obtained results, since $L_{cldice}$ and, especially, $L_{propdice}$, achieve higher sensitivity at the cost of decreased specificity, thus having lower accuracies. Regarding the proposed topological similarity indices, the runs minimizing $L_{propdice}$ were the best performing ones. One interesting finding was that the inclusion of the topological term $L_{topo}$ in $L_{propdice}$ also lead frequently to the increase of the clDice metric.

Concerning the BCE-based families, focusing the centrelines had only a negligible impact in the AUC. The remaining patterns follow the trends already discussed for the Dice family, with the proposed variants being once again the ones producing segmentations which are better topological-wise, according to the proposed similarity indices. An exception occurred in the Bcew family, where the proposed extension did not perform strictly better in the CORONARY database.

\begin{figure*}[t]
	\begin{minipage}{0.1\textwidth}
		\begin{center}
			Image
		\end{center}
	\end{minipage}
	\begin{minipage}{0.89\textwidth}
		\begin{center}
		\includegraphics[width=0.11\textwidth,height=0.11\textwidth]{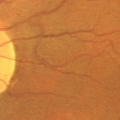}
		\includegraphics[width=0.11\textwidth,height=0.11\textwidth]{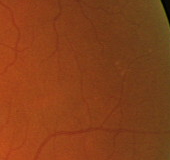}
		\includegraphics[width=0.11\textwidth,height=0.11\textwidth]{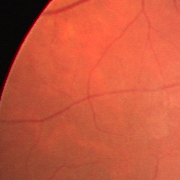}
		\includegraphics[width=0.11\textwidth,height=0.11\textwidth]{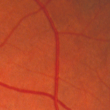}
		\includegraphics[width=0.11\textwidth,height=0.11\textwidth]{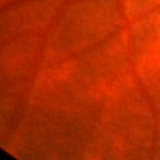}
		\includegraphics[width=0.11\textwidth,height=0.11\textwidth]{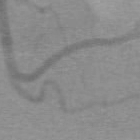}
		\includegraphics[width=0.11\textwidth,height=0.11\textwidth]{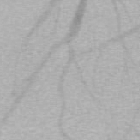}
		\includegraphics[width=0.11\textwidth,height=0.11\textwidth]{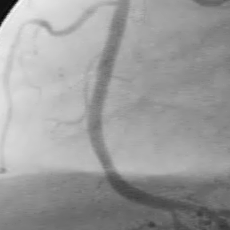}
		\end{center}
	\end{minipage}
	\vspace{1mm}

	\begin{minipage}{0.1\textwidth}
		\begin{center}
			GT
		\end{center}
	\end{minipage}
	\begin{minipage}{0.89\textwidth}
		\begin{center}
		\includegraphics[width=0.11\textwidth,height=0.11\textwidth]{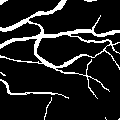}
		\includegraphics[width=0.11\textwidth,height=0.11\textwidth]{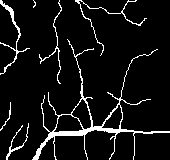}
		\includegraphics[width=0.11\textwidth,height=0.11\textwidth]{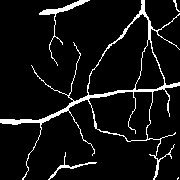}
		\includegraphics[width=0.11\textwidth,height=0.11\textwidth]{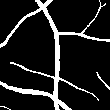}
		\includegraphics[width=0.11\textwidth,height=0.11\textwidth]{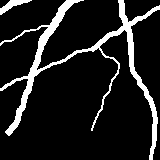}
		\includegraphics[width=0.11\textwidth,height=0.11\textwidth]{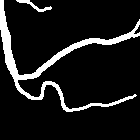}
		\includegraphics[width=0.11\textwidth,height=0.11\textwidth]{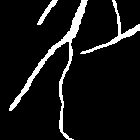}
		\includegraphics[width=0.11\textwidth,height=0.11\textwidth]{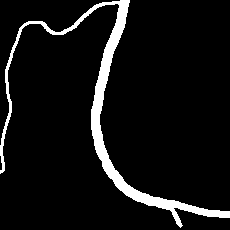}
		\end{center}
	\end{minipage}
	\vspace{1mm}
	
	\begin{minipage}{0.1\textwidth}
		\begin{center}
			Baseline
		\end{center}
	\end{minipage}
	\begin{minipage}{0.89\textwidth}
		\begin{center}
		\includegraphics[width=0.11\textwidth,height=0.11\textwidth]{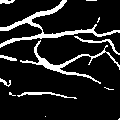}
		\includegraphics[width=0.11\textwidth,height=0.11\textwidth]{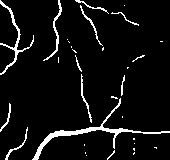}
		\includegraphics[width=0.11\textwidth,height=0.11\textwidth]{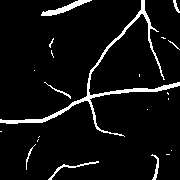}
		\includegraphics[width=0.11\textwidth,height=0.11\textwidth]{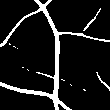}
		\includegraphics[width=0.11\textwidth,height=0.11\textwidth]{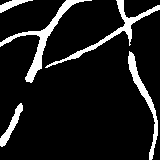}
		\includegraphics[width=0.11\textwidth,height=0.11\textwidth]{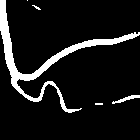}
		\includegraphics[width=0.11\textwidth,height=0.11\textwidth]{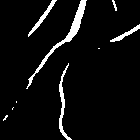}
		\includegraphics[width=0.11\textwidth,height=0.11\textwidth]{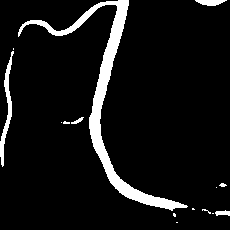}
		\end{center}
	\end{minipage}
	\vspace{1mm}
	
	\begin{minipage}{0.1\textwidth}
		\begin{center}
			Skeleton\\aware
		\end{center}
	\end{minipage}
	\begin{minipage}{0.89\textwidth}
		\begin{center}
		\includegraphics[width=0.11\textwidth,height=0.11\textwidth]{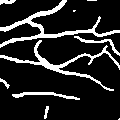}
		\includegraphics[width=0.11\textwidth,height=0.11\textwidth]{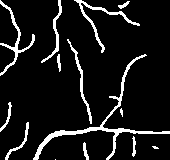}
		\includegraphics[width=0.11\textwidth,height=0.11\textwidth]{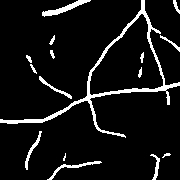}
		\includegraphics[width=0.11\textwidth,height=0.11\textwidth]{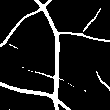}
		\includegraphics[width=0.11\textwidth,height=0.11\textwidth]{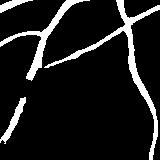}
		\includegraphics[width=0.11\textwidth,height=0.11\textwidth]{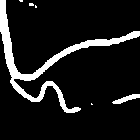}
		\includegraphics[width=0.11\textwidth,height=0.11\textwidth]{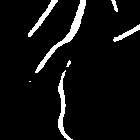}
		\includegraphics[width=0.11\textwidth,height=0.11\textwidth]{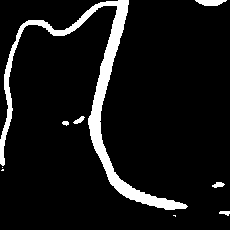}
		\end{center}
	\end{minipage}
	\vspace{1mm}
	
	\begin{minipage}{0.1\textwidth}
		\begin{center}
			Proposed
		\end{center}
	\end{minipage}
	\begin{minipage}{0.89\textwidth}
		\begin{center}
		\includegraphics[width=0.11\textwidth,height=0.11\textwidth]{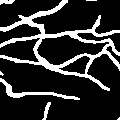}
		\includegraphics[width=0.11\textwidth,height=0.11\textwidth]{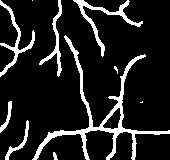}
		\includegraphics[width=0.11\textwidth,height=0.11\textwidth]{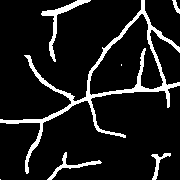}
		\includegraphics[width=0.11\textwidth,height=0.11\textwidth]{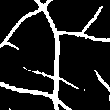}
		\includegraphics[width=0.11\textwidth,height=0.11\textwidth]{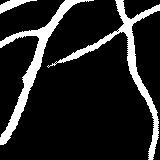}
		\includegraphics[width=0.11\textwidth,height=0.11\textwidth]{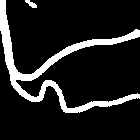}
		\includegraphics[width=0.11\textwidth,height=0.11\textwidth]{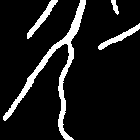}
		\includegraphics[width=0.11\textwidth,height=0.11\textwidth]{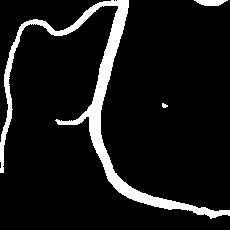}
		\end{center}
	\end{minipage}
\vspace{1mm}

	\caption{Example image patches, their ground truth, and segmentations achieved when minimizing the typical baseline loss functions ($L_{dice},L_{bce}$), the centreline-aware baseline loss functions ($L_{cldice},L_{clbce}$), and the extension of the latter with the proposed loss function ($L_{propdice}, L_{propbce}$). Segmentations from the best performing model in the particular dataset are shown. Coloured patches are from retinal images while the grayscale ones come from the coronary database.}
	\label{fig:topo_loss_visual_results}
\end{figure*}

The inferior performance of models belonging to the Dice and Bcew families in the CHASE dataset, when comparing with the ones from the Bceu family, was due to converging problems in some of the experimental runs. Having this in mind, and according to all of the obtained results, the best model to pick would possibly be the one trained with $L_{propbce}(\mathbf{P},\mathbf{Y};0.5,1,0.5,0.1,10)$. This model systematically performed better than the baselines topological-wise without disturbing significantly the pixel-wise metrics such as the AUC. Fig.~\ref{fig:topo_loss_visual_results} shows example segmentations that can be obtained when learning the model with the different loss functions. For each image, we show the outputs of the family of models that achieved the best performance (average between $\tilde{m}_H$ and $\tilde{m}_F$) on its dataset. For example, regarding the images coming from the CORONARY dataset, where the best performing model was the one trained with $L_{propdice}$, we show images for $L_{dice}, L_{cldice}$, and $L_{propdice}$. From the qualitative assessment of the visual results, it is possible to verify that the proposed loss term helped the model to produce segmentations with less errors that lead to disjoint trees, without introducing a significant number of errors that join distinct trees. The skeleton-aware baselines seem to have a tendency to overestimate the calibre of narrow segments, an effect that seems to increase even a bit more when being extended with the proposed loss term. This behaviour may introduce an error concerning the calibre estimation of narrow blood vessels and should be further investigated in the future.

\section{Conclusion and Future Work}

Deep learning has been pushing forward the performance levels of computer vision algorithms, and the blood vessel segmentation scenario is not an exception. Even then, a careful inspection of the outputs of deep networks allows to understand that they were not produced by a human expert. One type of errors that keeps affecting  these models is the non-detection of particularly challenging blood vessel segments, leading to disjoint trees and, therefore, having a great impact in the overall graph of the blood vessel tree. There is also the possibility of detecting false segments and joining distinct trees.

The lack of a single unified benchmark for assessing the topological properties of the segmented blood vessel trees certainly does not help raising awareness to this topic. To address this limitation, we designed a novel similarity index having properties which we deem necessary for assessing the topological coherence: i) to clearly highlight the errors that have impact on the vascular tree graphs; ii) to further penalize false negative segments inducing disjoint trees and false positive segments joining distinct trees; and iii) topological errors in the main vascular tree branches should have larger impact. 

Our experiments show that state-of-the-art losses promoting centreline consistency are not enough for improving topological coherence. A novel loss function focusing on the penalization of the errors described before in point ii), by means of a framework comprising the morphological closing operator, was proposed. The extension of centreline-aware losses with this term has helped models become more robust topological-wise.

We hope that this work brings more awareness to the need of, not only improving the resilience to these topological errors in future approaches, but also reporting how good a given methodology is topological-wise. We stress that, with the continuous evolution methodologies will keep facing, only reporting the typical metrics will be less sufficient as time goes by, since many properties of the produced segmentations will be overlooked.

\end{document}